\documentclass[twocolumn,superscriptaddress,floatfix,pra]{revtex4-2}
\usepackage[utf8]{inputenc}
\usepackage[T1]{fontenc}
\usepackage[english]{babel}
\usepackage{mathtools}
\usepackage{amsfonts}
\usepackage{mathptmx}
\usepackage{bm}
\usepackage{graphicx}
\usepackage{physics}
\usepackage[table,dvipsnames]{xcolor}
\usepackage[colorlinks=true,linkcolor=blue,citecolor=blue]{hyperref}
\usepackage{algorithm}
\usepackage{algpseudocode}
\usepackage{tikz}
\usetikzlibrary{positioning}
\usepackage{xcolor}

\begin{document}

\title{  
Quantum Information Transfer in Molecular Polaritonic Networks: \linebreak 
A Van Vleck Effective-Hamiltonian Approach
}

\author{Saad Yalouz}
\email{saad.yalouz@cnrs.fr}
\affiliation{
Institut de Physique et Chimie des Matériaux de Strasbourg,Université de Strasbourg, CNRS, UMR 7504, F-67000 Strasbourg, France
}

\author{Vincent Pouthier}
\affiliation{
Université Marie et Louis Pasteur, CNRS,
Institut UTINAM (UMR 6213), Equipe $\phi$th,
F-25000 Besan\c{c}on, France
}

\date{\today}
\begin{abstract}

In quantum information processing, achieving high-fidelity transfer of an arbitrary quantum state between spatially separated qubits is a central challenge for the development of reliable hardware architectures.
\linebreak
We address this question within a polaritonic framework by introducing a dedicated theoretical approach termed  \textit{Projected-Block Van Vleck Perturbation Theory} (PB-VVPT).
Based on quasidegenerate perturbation theory, this approach provides an effective-Hamiltonian description of the dynamics of a molecular network coupled to an optical cavity and local phonons, as modeled by a generalized Holstein--Tavis--Cummings Hamiltonian. 
\linebreak
Quantum information is encoded in a coherent superposition of exciton–cavity states, allowing light–matter degrees of freedom to act as qubit-state carriers, while phonons constitute a weakly coupled environment.  
\linebreak
In the nonadiabatic regime of weak exciton--phonon coupling, the method incorporates phonon-induced renormalization of polaritonic states and their dynamics while reducing the Hamiltonian to independent fixed-phonon-number sectors and coupled polaron--polariton subblocks. This structure provides efficient access to the coherence and population propagators governing quantum-state transfer without invoking a Markovian approximation. We apply the approach to a molecular wheel network and benchmark its predictions against exact Holstein--Tavis--Cummings dynamics, considering both molecule-to-molecule and molecule-to-cavity transfer.  
 We show that PB-VVPT accurately reproduces short- and long-time transfer dynamics, including thermal decoherence and fidelity revivals consistent with non-Markovian memory effects, with prediction errors remaining below $1\%$ over most of the simulated time intervals. 
This accuracy is achieved with an estimated diagonalization cost more than two orders of magnitude lower than that of the exact reference calculations.
Together, these results establish PB-VVPT as an accurate and computationally efficient approach to describing quantum information transfer in molecular polaritonic networks.

\end{abstract}

\maketitle

\section{Introduction }
\label{sec:introduction}

Polaritonic effects in optical cavities open new avenues for controlling quantum transport through light--matter coupling. When molecular excitations interact with cavity modes, hybrid light--matter states, known as \textit{polaritons}, emerge and introduce new transport channels that compete with direct intermolecular interactions~\cite{hertzog2019strong,tibben2023molecular,sandik2025cavity}.
The possibility of engineering these channels through the cavity environment has motivated extensive theoretical and experimental investigations into charge and energy transport. Theoretical studies have shown that cavity-mediated interactions can enhance exciton conductance~\cite{feist2015extraordinary,schachenmayer15}, charge transport~\cite{hagenmuller17}, and reshape vibrational energy transport~\cite{pouthier2026light}. In the presence of molecular vibrations, phonons can also facilitate cavity-mediated exciton transfer under specific coupling conditions~\cite{liu2019vibration}, while collective light--matter coupling can protect polaritonic coherence against phonon-induced decoherence~\cite{chng2024mechanism}. Experimentally, enhanced charge conductivity has been observed in organic semiconductors under strong light--matter coupling~\cite{orgiu2015conductivity}, while spatially resolved measurements have revealed exciton-polariton propagation over micrometre distances~\cite{rozenman18,hou2020ultralong}, with regimes ranging from enhanced diffusion to ultrafast ballistic motion~\cite{balasubrahmaniyam2023enhanced}.

Going beyond charge and energy transport, polaritonic systems have also attracted growing interest in the field of quantum information processing and, more specifically, for the realization of cavity-mediated information transfer~\cite{yang2022molecular,xiang2024molecular,modi2026molecular}. Several studies have explored this possibility in coupled-cavity arrays containing atomic or quantum-dot emitters~\cite{bose2007polaritonic,almeida2016staggered}, atomic cavity-QED systems~\cite{cirac1997quantum,ritter2012elementary}, and superconducting circuits~\cite{sillanpaa2007coherent,liu2017transmon,kurpiers2018deterministic}. Hybrid architectures have also been proposed to exchange information between a cavity photon bus and collective states of cold polar molecules~\cite{rabl2006hybrid}. In a related direction, experiments on molecular vibrational polaritons have demonstrated coherence transfer between coupled cavities~\cite{xiang2022coherence}.

Interestingly, while cavity-mediated information transfer has been investigated across various physical platforms, comparatively little attention has been devoted to its study in molecular excitonic networks accounting for the effects of a local phonon environment. 
Understanding how collective light--matter coupling and local molecular vibrations jointly shape quantum information transfer in these systems calls for an accurate and computationally efficient description of the underlying dynamics. This represents a central motivation of the present work.

We address this question within the framework of quantum-state transfer (QST), a fundamental protocol in quantum information processing aimed at transmitting an arbitrary quantum superposition, $\ket{\psi}=c_0\ket{0}+c_1\ket{1}$, between two spatially separated qubits with the highest possible fidelity~\cite{cirac1997quantum,Bose2003,Bose2007}. Unlike excitation transport alone, faithful QST requires preserving both the populations and the coherence of the initial quantum state. In previous work, we investigated how phonon-induced decoherence affects QST in molecular excitonic networks without optical cavities, with particular emphasis on the transfer of quantum coherence~\cite{pouthierPRB12,yalouz2017exciton,YalouzPouthier2016}. Here, we extend this description to polaritonic networks to investigate both population and coherence transfer, and to determine how their dynamics are modified by collective light--matter coupling in the presence of molecular vibrations.

To this end, we consider in this work a generalized Holstein--Tavis--Cummings (HTC) Hamiltonian describing molecular excitonic networks coupled to a common cavity mode and interacting with local molecular vibrations~\cite{cwik2014polariton,spano2015optical,wu2016polarons,herrera2017absorption}. Starting from this model, we develop a second-order effective-Hamiltonian approach based on quasidegenerate Van Vleck perturbation theory~\cite{vanvleck1929, shavitt1980} specifically adapted to the nonadiabatic regime of weak exciton--phonon coupling. The method, called \textit{Projected-Block Van Vleck Perturbation Theory} (PB-VVPT), incorporates the phonon-induced renormalization of polaritonic states while retaining
the residual couplings between states belonging to
selected polaron--polariton subblocks.
This treatment allows us to account explicitly for
degenerate and quasidegenerate states, whose coupled
dynamics cannot generally be described by independent
state corrections alone.
Together with the decomposition into fixed-phonon-number
sectors, this block structure reduces the numerical cost
and provides access to both the population and coherence
propagators governing QST, without invoking a Markovian
approximation. We analyze the computational cost of PB-VVPT and assess
its accuracy through systematic comparisons with numerically
exact HTC dynamics, distinguishing the effects of the
effective-Hamiltonian approximation from those of the
additional block filtering.
Applications to a molecular wheel network, considering
both site-to-site and site-to-cavity transfer, illustrate
how the method captures thermal decoherence, the
phonon-induced modification of optimal transfer conditions,
and long-time fidelity revivals.

The paper is organized as follows.
Section~\ref{sec:theory} introduces the generalized
HTC model and the QST framework.
In Sec.~\ref{sec:van_vleck}, we develop PB-VVPT
and derive the corresponding population and
coherence propagators.
Section~\ref{sec:numerical_evaluation} presents
their numerical implementation and analyzes
the computational cost.
Section~\ref{sec:numerical_results} reports
the numerical benchmarks and examines the
resulting transfer dynamics.
Finally, Sec.~\ref{sec:conclusion} summarizes
our results and discusses future perspectives.

\section{Polaritonic Model  and  Quantum-State Transfer Framework}
\label{sec:theory}
 
In this section, we introduce the theoretical framework
for studying QST in a molecular network embedded
in an optical cavity and locally coupled to vibrational
degrees of freedom. We first present the Hamiltonian describing
the exciton--cavity--phonon system.
We then formulate the QST protocol
and define the quantities used to characterize the transfer
of quantum information between molecular and photonic
degrees of freedom.

\subsection{ Generalized Holstein--Tavis--Cummings Hamiltonian}
\label{sec:htc}

\begin{figure*}
    \centering
    \includegraphics[width=16cm]{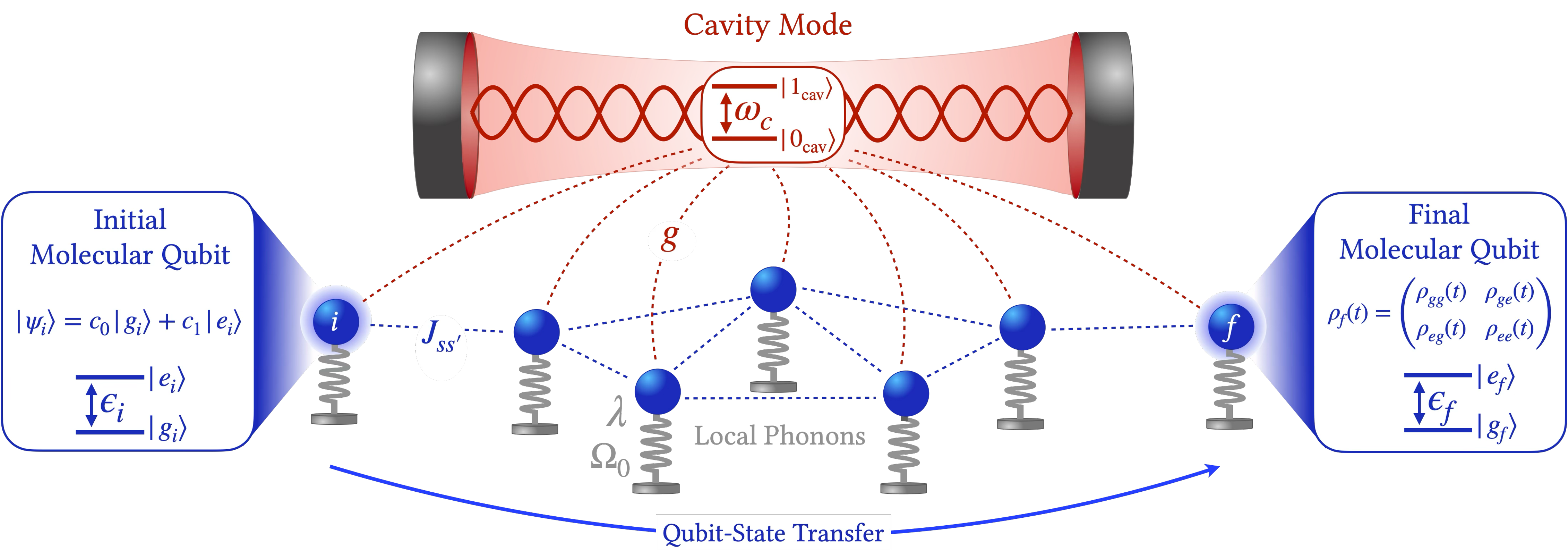}
   \caption{
Illustration of the quantum-state transfer in a cavity-coupled molecular network ruled by a generalized Holstein--Tavis--Cummings Hamiltonian.
Each molecular site is modeled as a two-level system (qubit), with ground and excited states
$\ket{g_s}$ and $\ket{e_s}$ separated by the excitation energy $\epsilon_s$.
The molecular excitations are coherently coupled through dipole-dipole interactions $J_{ss'}$
(blue dashed lines), while all sites interact with a common cavity mode of frequency
$\omega_c$ with uniform coupling strength $g$ (red dashed lines).
Each molecule is additionally coupled with strength $\lambda$ to a local phonon mode of
frequency $\Omega_0$ (gray oscillators).
In the QST protocol, an arbitrary qubit state 
$\ket{\psi_i}=c_0\ket{g_i}+c_1\ket{e_i}$ is initially encoded on site $i$ and subsequently
propagates through the coupled exciton--cavity--phonon system.
The
transferred information is characterized by the reduced density matrix
$\rho_f(t)$ of the qubit at the target site $f$ (see Eq.~\eqref{eq:final_qubit_density}).
Note that the cavity mode may also serve as the initial or target qubit, with logical states encoded in its vacuum and one-photon states.}
    \label{fig:model}
\end{figure*}

As illustrated in Fig.~\ref{fig:model},  we consider a network of $N$ identical molecular qubits  uniformly
coupled to a single optical cavity mode. Each molecule $s$ supports a
local ground state $\ket{g_s}$ and an excited state $\ket{e_s}$,
separated by an excitation energy $\epsilon_s$. The global ground
state of the exciton--cavity system is
$\ket{0_{\mathrm{exc-cav}}}
=\ket{0_{\mathrm{exc}}}\otimes\ket{0_{\mathrm{cav}}}$, where
$\ket{0_{\mathrm{exc}}}=\ket{g_1,\ldots,g_N}$ denotes the collective
molecular ground state and $\ket{0_{\mathrm{cav}}}$ the cavity
vacuum.
Here, we only consider the exciton--cavity Hilbert space restricted to the
zero- and one-excitation manifolds. The one-excitation manifold is
spanned by the molecular states
$\ket{s}=\ket{g_1,\ldots,e_s,\ldots,g_N}
\otimes\ket{0_{\mathrm{cav}}}$ and by the photonic state
$\ket{c}=\ket{0_{\mathrm{exc}}}\otimes\ket{1_{\mathrm{cav}}}$.
Within the rotating-wave approximation, the corresponding
exciton--cavity Hamiltonian takes the Tavis--Cummings
form~\cite{TavisCummings1968}, supplemented here by intermolecular
excitonic couplings (with $\hbar=1$),
\begin{equation}
\begin{aligned}
 H_{\mathrm{exc-cav}}
 ={}&
 \sum_{s=1}^{N}\epsilon_s\ketbra{s}{s}
 +\sum_{s>s'}J_{ss'}
 \left(\ketbra{s}{s'}+\ketbra{s'}{s}\right)\\
 &+\omega_c\ketbra{c}{c}
 +\sum_{s=1}^{N}g
 \left(\ketbra{c}{s}+\ketbra{s}{c}\right).
\end{aligned}
\label{eq:polaritonic_hamiltonian}
\end{equation}
Here, $J_{ss'}$ denotes the dipole--dipole coupling between molecular
sites, $\omega_c$ is the cavity frequency, and $g$ is the uniform
light--matter coupling.

Beyond the light--matter interaction, we also consider independent local phonon modes of frequency $\Omega_0$, described by the Hamiltonian
\begin{align}
 H_{\mathrm{ph}}
 &=\sum_{s=1}^{N}\Omega_0 b_s^\dagger b_s,
 \label{eq:phonon_hamiltonian}
\end{align}
where $b_s^\dagger$ and $b_s$ denote the creation and annihilation operators of the phonon mode associated with site $s$. 
Each molecular excitation is locally coupled to its corresponding phonon mode through a Holstein interaction of uniform strength $\lambda$ given by~\cite{Holstein1959}
\begin{align}
 H_{\mathrm{exc-ph}}
 &=\sum_{s=1}^{N}\lambda\ketbra{s}{s}
 \otimes(b_s^\dagger+b_s).
 \label{eq:exciton_phonon_coupling}
\end{align}
To highlight how the local phonons interact with the exciton--cavity system, we introduce the spectral decomposition of $H_{\mathrm{exc-cav}}$ within the one-excitation manifold:
\begin{equation}
 H_{\mathrm{exc-cav}}
 =
 \sum_{k=1}^{N+1}E_k\ketbra{\chi_k}{\chi_k},
 \label{eq:polariton_eigenproblem}
\end{equation}
where $\ket{\chi_k}$ and $E_k$ denote the ``\textit{unperturbed}'' polaritonic eigenstates and corresponding eigenenergies. In this polaritonic basis, the exciton--phonon interaction becomes
\begin{equation}
 H_{\mathrm{exc-ph}}
 =
 \sum_{s=1}^{N}\sum_{k,k'=1}^{N+1}
 M_{kk'}^{(s)}
 \ketbra{\chi_k}{\chi_{k'}}
 \otimes(b_s^\dagger+b_s),
 \label{eq:polariton_phonon_coupling}
\end{equation}
where we have introduced the phonon-mediated polaritonic coupling matrix elements
\begin{equation}
 M_{kk'}^{(s)}
 =
 \lambda\braket{\chi_k}{s}\braket{s}{\chi_{k'}},
 \label{eq:polariton_phonon_matrix_elements}
\end{equation}
with $s=1,\ldots,N$.
Comparing Eqs.~\eqref{eq:exciton_phonon_coupling} and
\eqref{eq:polariton_phonon_matrix_elements}, one sees that, although the
Holstein interaction is diagonal in the local molecular-site basis
$\lbrace\ket{s}\rbrace_{s=1}^{N}$, it gives rise to both diagonal and
off-diagonal contributions in the \textit{unperturbed} polaritonic basis
$\lbrace\ket{\chi_k}\rbrace_{k=1}^{N+1}$. The diagonal terms describe
the phonon dressing of individual polaritonic states, whereas the
off-diagonal terms generate phonon-assisted couplings between distinct
polaritonic states through the creation or annihilation of a phonon in
the local mode $s$. Both contributions are therefore directly governed
by the molecular character of the polaritonic states through the
amplitudes $\braket{s}{\chi_k}$ in
Eq.~\eqref{eq:polariton_phonon_matrix_elements}.

Putting all contributions together, the complete exciton--cavity--phonon system is described by the
generalized Holstein--Tavis--Cummings (HTC) Hamiltonian
\begin{equation}
 H=H_{\mathrm{exc-cav}}+H_{\mathrm{ph}}+H_{\mathrm{exc-ph}},
 \label{eq:total_hamiltonian}
\end{equation}
which combines intermolecular excitonic interactions, collective
light--matter coupling, and local phonon
dressing~\cite{cwik2014polariton,spano2015optical,wu2016polarons,herrera2017absorption}.
Importantly, note that under this Hamiltonian the total
exciton--photon excitation number is conserved, whereas the phonon
number is not.

\subsection{Quantum-State Transfer in the Polaritonic Framework}
\label{sec:qst}

Having introduced the HTC Hamiltonian that describes the light--matter system, we now turn to the description of a QST protocol within this polaritonic framework.
In the following, the excitonic network and the cavity are treated jointly as the quantum system that carries quantum information, while the phonons play the role of the perturbing external environment.
Following the framework introduced by
Bose~\cite{Bose2003,Bose2007}, we consider quantum information initially encoded in
the exciton--cavity subsystem through the two-level qubit superposition:
\begin{equation}
 \ket{\psi_i}
 =
 c_0\ket{0_{\mathrm{exc-cav}}}+c_1\ket{i} \quad
  \text{with} 
  \quad
 |c_0|^2+|c_1|^2=1.
 \label{eq:initial_qubit_state}
\end{equation}
Note here that the initial state $\ket{i}$ belongs to the one-excitation basis
$\{\ket{1},\ldots,\ket{N},\ket{c}\}$, so the encoded excitation
can be molecular or photonic in nature.
In the associated local basis
$\{\ket{0_{\mathrm{exc-cav}}},\ket{i}\}$, the density
matrix of the initial qubit carrying the information is given by
\begin{equation}
 \rho_i(0)
 =
 \ketbra{\psi_i}{\psi_i}
 =
 \begin{pmatrix}
 1- |c_1|^2 & c_0c_1^*\\
 c_1c_0^* & |c_1|^2
 \end{pmatrix}.
 \label{eq:initial_qubit_density}
\end{equation}
At the scale of the full system, we assume that the qubit-encoding process is fast compared with the characteristic phonon dynamics. Within this regime, the initial state preparation can be treated within the sudden approximation, such that the phonon state remains essentially unaffected during the encoding process. The initial total density operator can therefore be written in the factorized form
\begin{equation}
 \rho_{\mathrm{tot}}(0)
 =
 \ketbra{\psi_i}{\psi_i}\otimes\rho_{\mathrm{ph}},
 \label{eq:initial_total_density}
\end{equation}
where the phonon environment is assumed to be initially at thermal equilibrium (at temperature $T$)
\begin{equation}
 \rho_{\mathrm{ph}}
 =
 \frac{e^{-\beta H_{\mathrm{ph}}}}{Z_{\mathrm{ph}}},
 \label{eq:phonon_thermal_state}
\end{equation}
with $Z_{\mathrm{ph}}
 =
 \operatorname{Tr}_{\mathrm{ph}}
 \left[\exp(- \beta H_{\mathrm{ph}})\right]$ the phonon partition function and $\beta = 1/(k_{\rm B}T)$.

The next step of the QST protocol consists in letting the system's time evolution occur to propagate the information.
Such time evolution is encoded in the total density operator such as
\begin{equation}
    \rho_{\mathrm{tot}}(t) =e^{-iHt}\rho_{\mathrm{tot}}(0)e^{iHt},
\end{equation}
where $H$ is the generalized HTC Hamiltonian as previously defined in Eq.~\eqref{eq:total_hamiltonian}. 
Tracing out the phonon degrees of freedom first gives the reduced
density operator of the exciton--cavity subsystem
\begin{equation}
 \rho_{\mathrm{exc-cav}}(t)
 =
 \operatorname{Tr}_{\mathrm{ph}}
 \left[\rho_{\mathrm{tot}}(t)\right].
 \label{eq:reduced_density_operator}
\end{equation}
This reduced operator provides a description of how the quantum information spreads over the molecular and cavity degrees of freedom in the presence of the local phonon environment. 
Starting from this quantity, the question we then ask is how the initial information can be transferred towards a specific final qubit associated with the state
$\ket{f}\in\{\ket{1},\ldots,\ket{N},\ket{c}\}$ (see Fig.~\ref{fig:model}). Here again, note that this qubit can be either of molecular or photonic nature.
To characterize this transfer, we consider the reduced density operator associated with the corresponding local exciton--cavity qubit, denoted $\rho_f(t)$, obtained by tracing out all the remaining  degrees of freedom. The latter reads
\begin{equation}
 \rho_f(t)
 =
 \begin{pmatrix}
 1-|c_1|^2\mathcal G_{fi}^{\mathrm{pop}}(t)
 &
 c_0c_1^*\mathcal G_{fi}^{\mathrm{coh}}(t)^*
 \\
 c_1c_0^*\mathcal G_{fi}^{\mathrm{coh}}(t)
 &
 |c_1|^2\mathcal G_{fi}^{\mathrm{pop}}(t)
 \end{pmatrix}.
 \label{eq:final_qubit_density}
\end{equation}
As readily seen in Eq.~\eqref{eq:final_qubit_density}, the quantum information transfer connecting both qubits $i \rightarrow f $ is completely characterized by two quantities: \textit{(i)} the population propagator
$\mathcal G_{fi}^{\mathrm{pop}}(t)$, which measures the transfer of the
excited-state population, and \textit{{(ii)}} the coherence propagator
$\mathcal G_{fi}^{\mathrm{coh}}(t)$, which measures the transfer of the
vacuum--excitation coherence.
These propagators take the following closed forms
\begin{equation}
\begin{aligned}
 \mathcal G_{fi}^{\mathrm{coh}}(t)
 &=
 \operatorname{Tr}_{\mathrm{ph}}
 \left[
 \rho_{\mathrm{ph}}e^{iH_{\mathrm{ph}}t}
 \bra{f}e^{-iHt}\ket{i}
 \right],\\[1mm]
 \mathcal G_{fi}^{\mathrm{pop}}(t)
 &=
 \operatorname{Tr}_{\mathrm{ph}}
 \left[
 \rho_{\mathrm{ph}}
 \bra{i}e^{iHt}\ket{f}
 \bra{f}e^{-iHt}\ket{i}
 \right],
\end{aligned}
\label{eq:exact_propagators}
\end{equation}
where the partial trace over the phonon degrees of
freedom accounts for the statistical average over the initial phonon
state. 

Successful QST from qubit $i$ to qubit $f$ requires the final
state of the target qubit (Eq.~\eqref{eq:final_qubit_density}) to reproduce the initial
qubit state (Eq.~\eqref{eq:initial_qubit_density}) as closely as possible.
To quantify the quality of this transfer, we introduce the average transfer fidelity $\overline{\mathcal F}_{fi}(t)$, obtained by uniformly averaging over all pure initial qubit states on the Bloch sphere~\cite{Bose2003,Bose2007}.
As detailed in Appendix~\ref{app:fidelity}, the resulting (phase-corrected) average fidelity reads
\begin{equation}
 \overline{\mathcal F}_{fi}(t)
 =
 \frac12
 +\frac13\left|\mathcal G_{fi}^{\mathrm{coh}}(t)\right|
 +\frac16\mathcal G_{fi}^{\mathrm{pop}}(t).
 \label{eq:average_transfer_fidelity_main}
\end{equation}
This expression highlights the complementary roles of population
and coherence transfer in preserving quantum information.
In particular, perfect QST,
$\overline{\mathcal F}_{fi}(t)=1$, occurs when both
$\mathcal G_{fi}^{\mathrm{pop}}(t)=1$ and
$|\mathcal G_{fi}^{\mathrm{coh}}(t)|=1$.
Under these conditions, the target qubit fully recovers the
initial population and coherence magnitude, thereby reproducing
the initial qubit state up to a locally correctable phase~\cite{Bose2003,Bose2007}.

\section{Quantum-State Transfer via Projected Block Van Vleck Perturbation  Theory }
\label{sec:van_vleck} 

Assessing QST in the polaritonic framework considered here requires
evaluating both propagators introduced in
Eq.~\eqref{eq:exact_propagators}, which account for the influence
of the phonon environment on the exciton--cavity dynamics.
In the present work, we assume that the phonon environment
couples weakly to the exciton--cavity system and that
phonon-number-changing processes are energetically off-resonant.
Specifically, we consider the nonadiabatic weak-coupling regime \cite{silbey1980transport,sonnek1996coherence,pouthier2009narrow},
characterized by the following conditions
\begin{equation}
 \frac{\lambda}{\Omega_0}\ll1
 \quad\text{and}\quad
 \frac{\Delta E_{\mathrm{exc-cav}}}{\Omega_0}\ll1,
 \label{eq:perturbative_regime}
\end{equation}
where
$\Delta E_{\mathrm{exc-cav}}=\max_{k,k'}|E_k-E_{k'}|$
denotes the spectral width of the exciton--cavity Hamiltonian
introduced in Eq.~\eqref{eq:polariton_eigenproblem}.
In this regime, the excitonic interactions and light--matter
coupling are treated exactly through $H_{\mathrm{exc-cav}}$,
whereas the exciton--phonon interaction can be treated perturbatively.
Building on this perturbative context, we develop an
effective approach called PB-VVPT to evaluate polaritonic QST.
In the following, we will first describe the successive steps in the construction
of the PB-VVPT effective Hamiltonian, and then derive the corresponding
coherence and population propagators governing QST.

\begin{figure*}
    \centering
    \includegraphics[width=\textwidth]{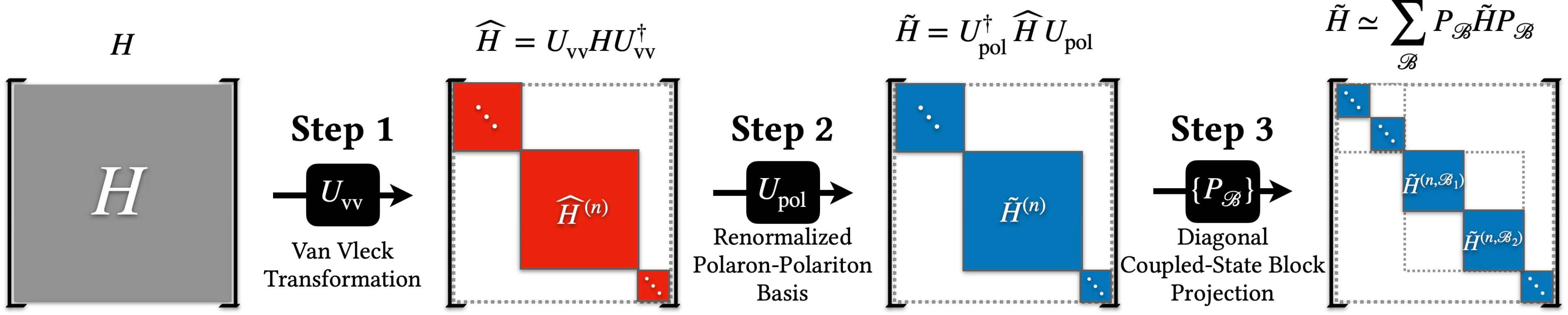}
\caption{
Projected-Block Van Vleck Perturbation Theory (PB-VVPT). 
Starting
from the original Holstein--Tavis--Cummings Hamiltonian, Step~1 applies the
Van~Vleck transformation $U_{\mathrm{vv}}$ to eliminate
phonon-number-changing couplings and generate independent fixed-$n$
sectors. In Step~2, the resulting Hamiltonian is expressed in the
renormalized polaron--polariton basis through $U_{\mathrm{pol}}$.
Finally, Step~3 identifies the $N_{\mathcal B}$ coupled-state blocks
from the remaining phonon-induced couplings and projects the dynamics
onto the corresponding subblocks.  
}
    \label{fig:effective_hamiltonian_scheme}
\end{figure*}

\subsection{Projected-Block Van~Vleck Perturbation Theory (PB-VVPT)}
\label{sec:effective_hamiltonian_construction} 

Starting from the perturbative conditions given in
Eq.~\eqref{eq:perturbative_regime}, one may partition the HTC Hamiltonian as follows:
\begin{equation}
 H=H_0+V,
 \label{eq:perturbative_partition}
\end{equation}
where
$H_0=H_{\mathrm{exc-cav}}+H_{\mathrm{ph}}$ is the unperturbed
exciton--cavity--phonon Hamiltonian, while
$V=H_{\mathrm{exc-ph}}$ describes the perturbative exciton--phonon
coupling.
An important consequence of the condition
$\Delta E_{\mathrm{exc-cav}}/\Omega_0\ll1$ is that the spectrum of
$H_0$ naturally separates into well-resolved manifolds associated with
a fixed total number of phonons. Each manifold contains the full
exciton--cavity spectrum and is separated from the neighboring ones by
an energy of order $\Omega_0$. By contrast, the perturbation $V$ couples neighboring phonon manifolds through single-phonon absorption-emission mechanisms (see Eq.~\eqref{eq:polariton_phonon_coupling}).
This perturbative coupling is entirely determined by the matrix elements $M_{kk'}^{(s)}$ introduced in
Eq.~\eqref{eq:polariton_phonon_matrix_elements}.

Starting from this partitioning, the derivation of the PB-VVPT effective 
Hamiltonian proceeds through three successive steps, summarized in
Fig.~\ref{fig:effective_hamiltonian_scheme}. First, a Van Vleck
transformation $U_{\mathrm{vv}}$ perturbatively removes the couplings
between different phonon-number manifolds, yielding independent
fixed-phonon-number sectors. 
Second, the resulting Hamiltonian is
expressed in a renormalized polaritonic basis through the unitary
transformation $U_{\mathrm{pol}}$. Finally, the remaining phonon-induced
couplings are analyzed and filtered to retain only groups of mutually coupled renormalized states.
\\

\paragraph*{\textbf{Step 1 -- Van Vleck Unitary Transformation:}}

Building on our previous works on exciton--phonon systems~\cite{YalouzPouthier2016,yalouzQInf17b,yalouz2017exciton}, we make use of quasi-degenerate Van Vleck perturbation theory to eliminate, up to second order in $\lambda$, the couplings between states belonging to different phonon-number sectors. To this end, we introduce the unitary transformation
$U_{\mathrm{vv}}=e^S$, with $S^\dagger=-S$, where the generator $S$ is determined perturbatively, order by order. The transformed Hamiltonian is defined as
\begin{equation}
 \widehat H = U_{\mathrm{vv}}HU_{\mathrm{vv}}^\dagger.
 \label{eq:van_vleck_transformation}
\end{equation}

By construction, and contrary to the original Hamiltonian $H$ of Eq.~\eqref{eq:total_hamiltonian}, for which the exciton--phonon interaction changes the total phonon number, $\widehat H$ conserves the latter up to the retained perturbative order. This effective Hamiltonian can therefore be decomposed into independent sectors of fixed total phonon number
\begin{equation}
 \widehat H
 =
 \bigoplus_{n=0}^{\infty}\widehat H^{(n)},
 \label{eq:van_vleck_phonon_blocks}
\end{equation}
where $\widehat H^{(n)}$ denotes the sector containing exactly $n$ phonons. This first block-diagonal structure is illustrated after the $U_{\mathrm{vv}}$ transformation in Fig.~\ref{fig:effective_hamiltonian_scheme}.

The resulting second-order effective Hamiltonian takes the following explicit form:
\begin{equation}
 \begin{aligned}
 \widehat H={}&H_{\mathrm{exc-cav}}+\delta H_{\mathrm{exc-cav}}\\
 +&H_{\mathrm{ph}} +\sum_{s,s'=1}^{N}\sum_{k,k'=1}^{N+1}
 \Lambda_{ss'}^{kk'}\ketbra{\chi_k}{\chi_{k'}}
 \otimes b_s^\dagger b_{s'},
 \end{aligned}
 \label{eq:effective_hamiltonian_second_order}
\end{equation}
whose derivation is presented in more detail in Appendix~\ref{app:van_vleck_expansion}. In Eq.~\eqref{eq:effective_hamiltonian_second_order}, the exciton--cavity correction $\delta H_{\mathrm{exc-cav}}$ is defined through the matrix elements
\begin{equation}
 \begin{aligned}
 &\bra{\chi_k}\delta H_{\mathrm{exc-cav}}\ket{\chi_{k'}}=\\
 &\frac{1}{2}\sum_{s=1}^{N}\sum_{k''=1}^{N+1}
 M^{(s)}_{kk''}M^{(s)}_{k''k'}
 \left(
 \frac{1}{E_k-E_{k''}-\Omega_0}
 +\frac{1}{E_{k'}-E_{k''}-\Omega_0}
 \right),
 \end{aligned}
 \label{eq:explicit_polariton_correction}
\end{equation}
and it has the effect of renormalizing the bare exciton--cavity energies and couplings.
Complementarily, the remaining phonon-number-conserving terms,
associated with the amplitudes
\begin{equation}
 \begin{aligned}
 \Lambda_{ss'}^{kk'}={}&\frac{1}{2}\sum_{k''=1}^{N+1}
 \Biggl[
 M^{(s)}_{kk''}M^{(s')}_{k''k'}
 \left(
 \frac{1}{E_k-E_{k''}+\Omega_0}
 +\frac{1}{E_{k'}-E_{k''}+\Omega_0}
 \right)\\
 &\quad+
 M^{(s')}_{kk''}M^{(s)}_{k''k'}
 \left(
 \frac{1}{E_k-E_{k''}-\Omega_0}
 +\frac{1}{E_{k'}-E_{k''}-\Omega_0}
 \right)
 \Biggr],
 \end{aligned}
 \label{eq:explicit_phonon_correction}
\end{equation}
encode the residual interaction between the polaritonic and phonon
degrees of freedom after the Van Vleck transformation.
For $s=s'$, this contribution describes the renormalization of the phonon frequency, whereas the terms
with $s\neq s'$ describe phonon transfers between different local modes.
\\

\paragraph*{\textbf{Step 2 -- Renormalized Polaron--Polariton Basis:}}
The Van Vleck effective Hamiltonian obtained in Step~1 contains two
different phonon-induced contributions: a renormalization
$\delta H_{\mathrm{exc-cav}}$ acting directly on the exciton--cavity
subsystem, and a residual coupling governed by the coefficients
$\Lambda_{ss'}^{kk'}$ that mix both exciton--cavity and phonon degrees of freedom.
To express these contributions in a more
convenient and compact form, we introduce a change of basis by
diagonalizing the exciton--cavity Hamiltonian contributions
\begin{equation}
 H_{\mathrm{exc-cav}}+\delta H_{\mathrm{exc-cav}}
 =
 \sum_{\mu=1}^{N+1}\widetilde E_\mu
 \ketbra{\widetilde\chi_\mu}{\widetilde\chi_\mu}.
 \label{eq:renormalized_polaritonic_diagonalization}
\end{equation}
The resulting energies $\widetilde E_\mu$ and states
$\ket{\widetilde\chi_\mu}$ therefore already include the second-order
phonon dressing of the original unperturbed polaritonic states. In the following,
we refer to the set
$\{\ket{\widetilde\chi_\mu}\}_{\mu=1}^{N+1}$ as the \textit{renormalized
polaron--polariton basis}. The latter is related to the original
unperturbed polaritonic basis through
\begin{equation}
 \ket{\widetilde\chi_\mu}
 =
 \sum_{k=1}^{N+1}
 \left(U_{\mathrm{pol}}\right)_{k\mu}\ket{\chi_k}, \quad  \text{with} \quad  
 \left(U_{\mathrm{pol}}\right)_{k\mu}
 =
 \braket{\chi_k}{\widetilde\chi_\mu}.
 \label{eq:renormalized_polaritonic_states}
\end{equation}

Within this \textit{renormalized polaron--polariton basis}, the transformed Van~Vleck effective
Hamiltonian, noted $ \widetilde H$,  reads
\begin{equation}
 \begin{aligned}
 \widetilde H
 &=
 U_{\mathrm{pol}}^\dagger\widehat H U_{\mathrm{pol}} \\
 &=
 \sum_{\mu=1}^{N+1}\widetilde E_\mu
 \ketbra{\widetilde\chi_\mu}{\widetilde\chi_\mu}\\
 &+ H_{\mathrm{ph}}+\sum_{s,s'=1}^{N}\sum_{\mu,\nu=1}^{N+1}
 \widetilde\Lambda_{ss'}^{\mu\nu}
 \ketbra{\widetilde\chi_\mu}{\widetilde\chi_\nu}
 \otimes b_s^\dagger b_{s'},
 \end{aligned}
 \label{eq:renormalized_polaritonic_hamiltonian_explicit}
\end{equation}
where the remaining phonon-induced couplings are given by
\begin{equation}
 \widetilde\Lambda_{ss'}^{\mu\nu}
 =
 \sum_{k,k'=1}^{N+1}
 \Lambda_{ss'}^{kk'}
 \left(U_{\mathrm{pol}}\right)_{k\mu}^{*}
 \left(U_{\mathrm{pol}}\right)_{k'\nu}.
 \label{eq:dressed_phonon_correction}
\end{equation}
Note that this transformation introduces no additional approximation.
As illustrated in Fig.~\ref{fig:effective_hamiltonian_scheme},
$U_{\mathrm{pol}}$ acts only on the exciton--cavity degrees of freedom
and simply changes the representation within each fixed-phonon-number
sector. It therefore preserves the block structure obtained in Step~1,
so that
\begin{equation}
 \widetilde H
 =
 \bigoplus_{n=0}^{\infty}\widetilde H^{(n)},
 \ \text{ with }\
 \widetilde H^{(n)}
 =
 U_{\mathrm{pol}}^\dagger
 \widehat H^{(n)}
 U_{\mathrm{pol}}.
 \label{eq:renormalized_polaritonic_phonon_blocks}
\end{equation}
In essence, Step~2 absorbs the phonon-induced renormalization
of the exciton--cavity subsystem into the definition of renormalized
polaron--polariton states, while the residual couplings to the phonon
degrees of freedom remain entirely encoded in
$\widetilde\Lambda_{ss'}^{\mu\nu}$.
These couplings generally mix different renormalized
polaron--polariton states within each fixed-phonon-number sector
and constitute the remaining information to be addressed, which is the objective of the
final step presented below.
\\

\paragraph*{\textbf{Step 3 -- Final Projection onto Coupled-State Sub-Blocks:}}

Starting from the renormalized polaron--polariton basis introduced in
Step~2, we now propose a simplified treatment of the residual
phonon-induced couplings $\widetilde\Lambda_{ss'}^{\mu\nu}$. The idea
is to compare their strength with the corresponding energy separation
between renormalized polaron--polariton states in order to identify which states
must remain coupled within the effective description. To this end, for
each pair of states with a nonvanishing energy separation $|\widetilde E_\mu-\widetilde E_\nu|$,  we introduce
the dimensionless ratio
\begin{equation}
 R_{\mu\nu}
 =
 \frac{\|\widetilde\Lambda^{\mu\nu}\|_{\mathrm F}}
 {|\widetilde E_\mu-\widetilde E_\nu|}
 \quad \text{with} \quad
 \|\widetilde\Lambda^{\mu\nu}\|_{\mathrm F}
 =
 \left(
 \sum_{s,s'=1}^{N}
 |\widetilde\Lambda_{ss'}^{\mu\nu}|^2
 \right)^{1/2}.
 \label{eq:block_coupling_ratio}
\end{equation}
The ratio $R_{\mu\nu}$ therefore measures the relative importance
of the residual phonon-induced coupling between two renormalized polaron--polariton
states with respect to their energy separation.
In practice, we consider two states as effectively ``coupled''
whenever $R_{\mu\nu}\geq R_{\mathrm c}$, where $R_{\mathrm c}$
is a numerical threshold to be specified.
This criterion naturally identifies coupled quasidegenerate states,
but also allows energetically separated states to be considered
coupled whenever their residual phonon-induced interaction
is sufficiently strong. Note that exactly degenerate states are systematically retained within the same block, independently of this criterion.
Grouping states via this approach
defines a partition of the exciton--cavity Hilbert space into
$N_{\mathcal B}$ disjoint coupled-state blocks $\mathcal B$.
Since the resulting blocks span the complete exciton--cavity
space, their dimensions satisfy
\begin{equation}
 \sum_{\mathcal B}\dim(\mathcal B)=N+1.
 \label{eq:coupled_state_block_partition}
\end{equation}
where the sum  $\sum_{\mathcal B}$ runs over all $N_{\cal B}$ blocks.
For each coupled-state block $\mathcal B$, we then introduce the
projector on the coupled renormalized polaron--polariton states
\begin{equation}
 P_{\mathcal B}
 =
 \sum_{\mu\in\mathcal B}
 \ketbra{\widetilde\chi_\mu}{\widetilde\chi_\mu},
 \label{eq:coupled_state_projector}
\end{equation}
 where $\mu\in\mathcal B$ denotes the indices of the renormalized
polaron--polariton states belonging to block $\mathcal B$.
Using these projectors, the Van~Vleck effective Hamiltonian given in Eq.~\eqref{eq:renormalized_polaritonic_hamiltonian_explicit} can be
exactly decomposed into intra- and interblock contributions as
\begin{equation}
 \widetilde H
 =
 \sum_{\mathcal B_1}
 P_{\mathcal B_1}\widetilde H P_{\mathcal B_1}
 +
 \sum_{\mathcal B_1\neq\mathcal B_2}
 P_{\mathcal B_1}\widetilde H P_{\mathcal B_2}.
 \label{eq:interblock_hamiltonian_decomposition}
\end{equation}

From this decomposition, the final form of the PB-VVPT Hamiltonian is obtained by introducing the
\textit{coupled-state block-diagonal approximation}, which consists in
neglecting the residual couplings between distinct blocks while retaining
all couplings within each block. The effective Hamiltonian then reduces to
\begin{equation}
 \widetilde H
 \simeq
 \sum_{\mathcal B}
 P_{\mathcal B}\widetilde H P_{\mathcal B}
 =
 \bigoplus_{\mathcal B}\widetilde H^{(\mathcal B)}.
 \label{eq:block_effective_hamiltonian}
\end{equation}
In terms of the residual phonon-induced couplings introduced in
Step~2, this approximation simply amounts to imposing
\begin{equation}
 \widetilde\Lambda_{ss'}^{\mu\nu}
 \longrightarrow
 \begin{cases}
  \widetilde\Lambda_{ss'}^{\mu\nu}
  & \text{for } \mu,\nu\in\mathcal B_1,\\
  0
  & \text{for } \mu\in\mathcal B_1,\ \nu\in\mathcal B_2,
    \text{ and } \mathcal B_1\neq\mathcal B_2.
 \end{cases}
 \label{eq:interblock_decoupling}
\end{equation}
In essence, all residual phonon-induced couplings between states belonging to the
same block $\mathcal{B}_1$ are retained, whereas all those connecting
different blocks $\mathcal{B}_1 \neq \mathcal{B}_2$ are discarded.

As illustrated in the last stage of 
Fig.~\ref{fig:effective_hamiltonian_scheme}, since $P_{\mathcal B}$ acts only on the exciton--cavity
degrees of freedom, this additional decomposition preserves the
fixed-phonon-number structure obtained in Step~1 and 2. Within each
$n$-phonon sector, we can therefore define
\begin{equation}
 \widetilde H^{(n,\mathcal B)}
 =
 P_{\mathcal B}\widetilde H^{(n)}P_{\mathcal B},
 \label{eq:effective_hamiltonian_block_n}
\end{equation}
such that the final effective Hamiltonian takes the doubly
block-diagonal form
\begin{equation}
 \widetilde H
 \simeq
 \bigoplus_{n=0}^{\infty}
 \bigoplus_{\mathcal B  } 
 \widetilde H^{(n,\mathcal B)}.
 \label{eq:double_block_structure}
\end{equation}
The effective exciton--cavity--phonon dynamics within the PB-VVPT approach is thus resolved into independent
$(n,\mathcal B)$ subspaces, where $n$ fixes the total phonon number
and $\mathcal B$ identifies the corresponding coupled
polaron--polariton subspace.

\subsection{Quantum-State Transfer \textit{via} PB-VVPT: \linebreak Explicit Form of the Coherence and Population Propagators  }
\label{sec:effective_propagators}

We now use the PB-VVPT strategy introduced above to derive explicit
expressions for the coherence and population propagators governing the
QST process. Starting from Step~1, the original Hamiltonian is first
transformed through the Van~Vleck unitary $U_{\mathrm{vv}}$ into the
second-order effective Hamiltonian $\widehat H$. The propagators, initially defined in Eq.~\eqref{eq:exact_propagators}, can
therefore be rewritten as
\begin{equation}
 \begin{aligned}
 \mathcal G_{fi}^{\mathrm{coh}}(t)
 &\simeq
 \operatorname{Tr}_{\mathrm{ph}}\Bigl[
 \rho_{\mathrm{ph}}e^{iH_{\mathrm{ph}}t}
 \bra{f}U_{\mathrm{vv}}^\dagger e^{-i\widehat Ht}U_{\mathrm{vv}}\ket{i}
 \Bigr],\\
 \mathcal G_{fi}^{\mathrm{pop}}(t)
 &\simeq
 \operatorname{Tr}_{\mathrm{ph}}\Bigl[
 \rho_{\mathrm{ph}}
 \bra{i}U_{\mathrm{vv}}^\dagger e^{i\widehat Ht}U_{\mathrm{vv}}\ket{f}
 \bra{f}U_{\mathrm{vv}}^\dagger e^{-i\widehat Ht}U_{\mathrm{vv}}\ket{i}
 \Bigr].
 \end{aligned}
 \label{eq:transformed_effective_propagators}
\end{equation}
As can be seen from this expression, a consistent second-order
Van Vleck treatment requires  not only the
corrections contained in $\widehat H$, but also the perturbative
reconstruction of the physical initial and final states through
$U_{\mathrm{vv}}$ and $U_{\mathrm{vv}}^\dagger$.
These additional terms account for the exciton--phonon dressing
of the original unperturbed exciton-cavity states and contribute at the same perturbative order.
However, previous studies by the present authors on
exciton--phonon coherence dynamics have shown that, in the regimes
investigated, these state-reconstruction corrections only weakly
affect the dynamics, which is already accurately reproduced by
retaining the corrections in the effective Hamiltonian while
treating the transformed states at zeroth
order~\cite{yalouz2017exciton,yalouzQInf17b,pouthier2011excitonic,pouthierPRB12}.
Motivated by this, we choose to adopt here the  approximation $U_{\mathrm{vv}}\simeq\mathbb I$ when reconstructing
the physical states, so that
\begin{equation}
 e^{-iHt}
 \simeq
 U_{\mathrm{vv}}^\dagger e^{-i\widehat Ht}U_{\mathrm{vv}}
 \xrightarrow{\,U_{\mathrm{vv}}\simeq\mathbb I\,}
 e^{-i\widehat Ht}.
 \label{eq:effective_propagator_approximation}
\end{equation}
Importantly, this approximation leaves the second-order construction
of $\widehat H$ unchanged, and only the reconstruction of the physical
states is truncated at zeroth order.

We next apply Steps~2 and~3 of the PB-VVPT construction to the
effective propagator $e^{-i\widehat Ht}$. The Hamiltonian is first
expressed in the renormalized polaron--polariton basis,
$\widehat H\rightarrow\widetilde H$, and then projected onto the
coupled-state subblocks introduced in Step~3. Combined with the
fixed-phonon-number decomposition obtained in Step~1, this leads to the
doubly block-diagonal structure given in Eq.~\eqref{eq:double_block_structure}.
Within this framework, the
propagators take the final simplified form
\begin{equation}
 \begin{aligned}
 \mathcal G_{fi}^{\mathrm{coh}}(t)
 &\simeq
 \sum_{n=0}^{\infty}
 \frac{e^{-\beta n\Omega_0}}{Z_{\mathrm{ph}}}e^{in\Omega_0t}
 \sum_{\mathcal B} 
 \operatorname{Tr}_{\mathrm{ph}}^{(n)}
 \Bigl[
 \bra{f}P_{\mathcal B} 
 e^{-i\widetilde H^{(n,\mathcal B)}t}
 P_{\mathcal B}\ket{i}
 \Bigr],\\
 \mathcal G_{fi}^{\mathrm{pop}}(t)
 &\simeq
 \sum_{n=0}^{\infty}
 \frac{e^{-\beta n\Omega_0}}{Z_{\mathrm{ph}}}
 \sum_{\mathcal B_1,\mathcal B_2} 
 \operatorname{Tr}_{\mathrm{ph}}^{(n)}
 \Bigl[
 \bra{i}P_{\mathcal B_2}
 e^{i\widetilde H^{(n,\mathcal B_2)}t}
 P_{\mathcal B_2}\ket{f}\\
 &\hspace{3.5cm}\times
 \bra{f}P_{\mathcal B_1}
 e^{-i\widetilde H^{(n,\mathcal B_1)}t}
 P_{\mathcal B_1}\ket{i}
 \Bigr].
 \end{aligned}
 \label{eq:block_population_fixed_n}
\end{equation}
Here, $\operatorname{Tr}_{\mathrm{ph}}^{(n)}[\cdots]$ denotes the
phonon trace restricted to the sector containing exactly $n$ phonons.
The factors $e^{-\beta n\Omega_0}$ and $e^{in\Omega_0t}$ appearing
above follow directly from the thermal and free evolution of the
phonons within each fixed-$n$ sector.

The final expressions for the PB-VVPT propagators given in
Eq.~\eqref{eq:block_population_fixed_n} provide all the  information
needed to fully characterize the final reduced density matrix in
Eq.~\eqref{eq:final_qubit_density}, and the
transfer fidelity of a QST protocol as defined in
Eq.~\eqref{eq:average_transfer_fidelity_main}.
In the particular limit where the renormalized polaron--polariton
states are all nondegenerate and effectively decoupled, each state
defines an independent block with $\dim(\mathcal B)=1$ (and $N_{\mathcal B}=N+1$ independent subblocks).
In this very special case, we can show that the above expressions simplify further and admit closed analytical forms.
We refer the interested reader to
Appendix~\ref{app:nondegenerate_reduction} for their explicit
expressions and derivation.
The numerical formulation introduced hereafter is general and does not rely on
this simplification, and it will allow treatment of
coupled-state blocks of arbitrary dimension.

\section{Numerical Implementation in  a Finite Many-Body Basis}
\label{sec:numerical_evaluation}

Having derived the PB-VVPT propagators, we now present
a numerical strategy for evaluating them through an exact
matrix representation within a finite many-body basis.
The same representation will also be used to evaluate the
original propagators under the full HTC Hamiltonian,
providing a reference for benchmarking the PB-VVPT predictions.

\subsection{Truncation of the Phonon Hilbert Space}
A numerical matrix representation of the coupled
exciton--cavity--phonon system requires truncating the
formally infinite phonon Fock space.
Here, we impose a maximum total phonon number
$N_{\mathrm{ph}}^{\max}$ and retain the occupation-number
states $\ket{\mathbf n}=\ket{n_1,\ldots,n_N}$ such that
$\sum_{s=1}^{N}n_s\leq N_{\mathrm{ph}}^{\max}$,
where $n_s$ denotes the occupation of the local phonon mode $s$.
The resulting truncated Fock space decomposes then as
\begin{equation}
 \mathcal H_{\mathrm{ph}}^{\mathrm{trunc}}
 =
 \bigoplus_{n=0}^{N_{\mathrm{ph}}^{\max}}
 \mathcal H_{\mathrm{ph}}^{(n)},
\end{equation}
where each subspace $\mathcal H_{\mathrm{ph}}^{(n)}$
is spanned by the occupation-number states with a fixed
total phonon number $\sum_{s=1}^{N}n_s=n$.
Each of these subspaces has a dimension
\begin{equation}
 D^{(n)} = \binom{N+n-1}{n},
 \label{eq:fixed_n_phonon_dimension}
\end{equation}
so that the total dimension of the truncated phonon Fock space reads $\dim(\mathcal H_{\mathrm{ph}}^{\mathrm{trunc}})  =
 \sum_{n=0}^{N_{\mathrm{ph}}^{\max}}D^{(n)} $.
This construction applies to both the PB-VVPT and HTC
Hamiltonians, irrespective of whether the dynamics conserves
the total phonon number.
In practice, the choice of the cutoff $N_{\mathrm{ph}}^{\max}$ must be sufficiently large to make  both  the thermal weight of the omitted sectors negligible and to ensure convergence of the HTC dynamics over the time interval considered (see Appendix~\ref{app:convergence} for discussion on this point). For a consistent comparison, the same truncated phonon space and normalized truncated thermal state will be used later on in both approaches.

\subsection{Explicit Matrix Representation of the  Propagators}
\label{sec:many_body_numerical_representations}
\subsubsection{Effective approach using the PB-VVPT Hamiltonian}

Within the PB-VVPT approach, neglecting the residual interblock
couplings allows us to construct the matrix representation
of the effective Hamiltonian separately within each independent
$(n,\mathcal B)$ subspace.
For a fixed total phonon number $n$ and coupled-state block
$\mathcal B$, we use the many-body basis
$\ket{\widetilde\chi_\mu,\mathbf n}
=\ket{\widetilde\chi_\mu}\otimes\ket{\mathbf n}$,
where $\mu\in\mathcal B$ labels the renormalized polaron--polariton
states belonging to block $\mathcal B$, and
$\ket{\mathbf n}\in\mathcal H_{\mathrm{ph}}^{(n)}$ runs over
the occupation-number states with total phonon number $n$.
Here and below, sums over phonon states are understood to run
over the occupation-number basis states of the indicated subspace.
In this basis, the corresponding PB-VVPT Hamiltonian subblock
$\widetilde H^{(n,\mathcal B)}$ is represented by a matrix
of dimension
\begin{equation}
 D^{(n,\mathcal B)} = \dim(\mathcal B)D^{(n)} .
\end{equation}
Diagonalizing the full PB-VVPT Hamiltonian amounts to solving
a  series of independent eigenvalue problems within each  $(n,\mathcal B)$ subspace:
\begin{equation}
 \widetilde H^{(n,\mathcal B)}
 \ket{\Phi_{\alpha}^{(n,\mathcal B)}}
 =
 \mathcal E_{\alpha}^{(n,\mathcal B)}
 \ket{\Phi_{\alpha}^{(n,\mathcal B)}},
 \label{eq:effective_block_diagonalization}
\end{equation}
where $\alpha=1,\ldots,D^{(n,\mathcal B)}$ labels the
many-body eigenstates within the $(n,\mathcal B)$ subspace.
Using the resulting eigenstates and eigenenergies to evaluate
the PB-VVPT propagators of
Eq.~\eqref{eq:block_population_fixed_n}, we obtain
\begin{widetext}
\begin{equation}
\begin{aligned}
\mathcal G_{fi}^{\mathrm{coh}}(t)
&\simeq
\sum_{n=0}^{N_{\mathrm{ph}}^{\max}}
\frac{e^{-\beta n\Omega_0}}{Z_{\mathrm{ph}}^{\mathrm{trunc}}}
e^{in\Omega_0t}
\sum_{\mathcal B}
\sum_{\mu,\nu\in\mathcal B}
\braket{f}{\widetilde\chi_\mu}
\braket{\widetilde\chi_\nu}{i}
\sum_{\alpha=1}^{D^{(n,\mathcal B)}}
e^{-i\mathcal E_{\alpha}^{(n,\mathcal B)}t}
\sum_{\ket{\mathbf n}\in\mathcal H_{\mathrm{ph}}^{(n)}}
\braket{\widetilde\chi_\mu,\mathbf n}
        {\Phi_{\alpha}^{(n,\mathcal B)}}
\braket{\Phi_{\alpha}^{(n,\mathcal B)}}
        {\widetilde\chi_\nu,\mathbf n},
\\[1ex]
\mathcal G_{fi}^{\mathrm{pop}}(t)
&\simeq
\sum_{n=0}^{N_{\mathrm{ph}}^{\max}}
\frac{e^{-\beta n\Omega_0}}{Z_{\mathrm{ph}}^{\mathrm{trunc}}}
\sum_{\mathcal B_1,\mathcal B_2}
\sum_{\mu_1,\nu_1\in\mathcal B_1}
\braket{f}{\widetilde\chi_{\mu_1}}
\braket{\widetilde\chi_{\nu_1}}{i}
\sum_{\mu_2,\nu_2\in\mathcal B_2}
\braket{i}{\widetilde\chi_{\nu_2}}
\braket{\widetilde\chi_{\mu_2}}{f}
\sum_{\alpha=1}^{D^{(n,\mathcal B_1)}}
\sum_{\alpha'=1}^{D^{(n,\mathcal B_2)}}
e^{-i(\mathcal E_{\alpha}^{(n,\mathcal B_1)}
      -\mathcal E_{\alpha'}^{(n,\mathcal B_2)})t}
\\
&\qquad\times
\sum_{\ket{\mathbf n}\in\mathcal H_{\mathrm{ph}}^{(n)}}
\braket{\Phi_{\alpha}^{(n,\mathcal B_1)}}
        {\widetilde\chi_{\nu_1},\mathbf n}
\braket{\widetilde\chi_{\nu_2},\mathbf n}
        {\Phi_{\alpha'}^{(n,\mathcal B_2)}}
\sum_{\ket{\mathbf m}\in\mathcal H_{\mathrm{ph}}^{(n)}}
\braket{\widetilde\chi_{\mu_1},\mathbf m}
        {\Phi_{\alpha}^{(n,\mathcal B_1)}}
\braket{\Phi_{\alpha'}^{(n,\mathcal B_2)}}
        {\widetilde\chi_{\mu_2},\mathbf m}.
\end{aligned}
\label{eq:numerical_effective_propagators}
\end{equation}
\end{widetext}
The thermal weights in
Eq.~\eqref{eq:numerical_effective_propagators} are normalized
within the truncated phonon space through the partition function
\begin{equation}
 Z_{\mathrm{ph}}^{\mathrm{trunc}}
 =
 \sum_{n=0}^{N_{\mathrm{ph}}^{\max}}
 D^{(n)}  e^{-\beta n\Omega_0},
 \label{eq:Z_part}
\end{equation}
where $D^{(n)} $, as defined in Eq.~\eqref{eq:fixed_n_phonon_dimension},  counts all the occupation-number states sharing
the same phonon energy $n\Omega_0$.

\subsubsection{Exact approach using the HTC Hamiltonian}

For comparison, we will also consider the direct numerical treatment
of the HTC Hamiltonian within the same truncated phonon space.
Unlike the PB-VVPT Hamiltonian, the HTC Hamiltonian does not
conserve the total phonon number, since the exciton--phonon
interaction couples neighboring phonon-number sectors (see Eq.~\eqref{eq:exciton_phonon_coupling}).
Its matrix representation must therefore include all retained
sectors simultaneously.
We use the many-body basis
$\ket{a,\mathbf n}=\ket{a}\otimes\ket{\mathbf n}$,
where $a\in\{1,\ldots,N,c\}$ labels the molecular excitations
and the cavity state, and
$\ket{\mathbf n}\in\mathcal H_{\mathrm{ph}}^{\mathrm{trunc}}$
runs over the retained phonon occupation-number states.
In this basis, the HTC Hamiltonian is represented by a single
$D\times D$ matrix, with
\begin{equation}
 D=(N+1)\sum_{n=0}^{N_{\mathrm{ph}}^{\max}}D^{(n)} .
 \label{eq:dim_HTC}
\end{equation}
The corresponding eigenvalue problem reads
\begin{equation}
 H\ket{\Psi_\alpha}
 =
 \mathcal E_\alpha\ket{\Psi_\alpha},
 \label{eq:exact_many_body_diagonalization}
\end{equation}
where $H$ denotes the HTC Hamiltonian projected onto the
truncated many-body space, and $\alpha=1,\ldots,D$ labels
its eigenstates $\ket{\Psi_\alpha}$ and eigenenergies
$\mathcal E_\alpha$.
Using these eigenstates and eigenenergies, the coherence
and population propagators can be evaluated directly from Eq.~\eqref{eq:exact_propagators} as
\begin{equation}
\begin{aligned}
\mathcal G_{fi}^{\mathrm{coh}}(t)
&=
\sum_{\ket{\mathbf n}\in
\mathcal H_{\mathrm{ph}}^{\mathrm{trunc}}}
\frac{e^{-\beta\varepsilon_{\mathbf n}}}
{Z_{\mathrm{ph}}^{\mathrm{trunc}}}
\sum_{\alpha=1}^{D}
e^{-i(\mathcal E_\alpha-\varepsilon_{\mathbf n})t}
\braket{f,\mathbf n}{\Psi_\alpha}
\braket{\Psi_\alpha}{i,\mathbf n},
\\
\mathcal G_{fi}^{\mathrm{pop}}(t)
&=
\sum_{\ket{\mathbf n},\ket{\mathbf m}\in
\mathcal H_{\mathrm{ph}}^{\mathrm{trunc}}}
\frac{e^{-\beta\varepsilon_{\mathbf n}}}
{Z_{\mathrm{ph}}^{\mathrm{trunc}}}
\sum_{\alpha,\alpha'=1}^{D}
e^{-i(\mathcal E_\alpha-\mathcal E_{\alpha'})t}
\\
&\qquad\times
\braket{f,\mathbf m}{\Psi_\alpha}
\braket{\Psi_\alpha}{i,\mathbf n}
\braket{i,\mathbf n}{\Psi_{\alpha'}}
\braket{\Psi_{\alpha'}}{f,\mathbf m}.
\end{aligned}
\label{eq:numerical_exact_propagators}
\end{equation}
Here, $\varepsilon_{\mathbf n}=\Omega_0\sum_s n_s$, and
the thermal weights are normalized using the same truncated
partition function $Z_{\mathrm{ph}}^{\mathrm{trunc}}$
as in the PB-VVPT calculation (see Eq.~\eqref{eq:Z_part}).
These expressions are numerically exact within the truncated
many-body Hilbert space.

 \subsection{Computational-Cost Comparison:\linebreak
PB-VVPT Hamiltonian  \textit{vs.} Exact HTC Hamiltonian}
\label{sec:computational_cost}

Using the same phonon cutoff and thermal normalization enables a direct comparison between the PB-VVPT method and the exact HTC reference dynamics for polaritonic QST.
Their main numerical difference lies in the structure of the corresponding many-body Hamiltonians to be treated.
Within PB-VVPT, the dynamics is resolved into independent $(n,\mathcal B)$ subspaces, whereas the full HTC Hamiltonian couples neighboring phonon-number sectors, requiring all retained sectors to be treated jointly.
To quantify the computational savings resulting from this block structure, we compare below the costs of solving the eigenvalue problems defined in Eqs.~\eqref{eq:effective_block_diagonalization} and~\eqref{eq:exact_many_body_diagonalization}, assuming dense diagonalization.

For the full HTC calculation, the Hamiltonian is represented in a many-body space of dimension $D$, given in Eq.~\eqref{eq:dim_HTC}.
The computational cost of dense diagonalization therefore scales as
\begin{equation}
 C_{\mathrm{HTC}}
 =
 \mathcal O\left[
 (N+1)^3\bigg(\sum_{n=0}^{N_{\mathrm{ph}}^{\max}}D^{(n)}  \bigg)^3
 \right].
 \label{eq:cost_exact_preparation}
\end{equation}

Within PB-VVPT, the successive reductions introduced above lead
instead to independent matrices of dimension
$D^{(n,\mathcal B)}=\dim(\mathcal B)D^{(n)} $.
The total cost associated with the dense diagonalization of all
$(n,\mathcal B)$ subspaces therefore scales as
\begin{equation}
C_{\text{\normalfont PB-VVPT}}
 =
 \mathcal O\left[
 \sum_{\mathcal B}\dim(\mathcal B)^3
 \sum_{n=0}^{N_{\mathrm{ph}}^{\max}}(D^{(n)}) ^3
 \right].
 \label{eq:cost_effective_preparation}
\end{equation}
This expression directly reflects the two successive reductions
underlying the PB-VVPT strategy: the decomposition of the phonon
Hilbert space into independent fixed-$n$ sectors, followed by
the decomposition of the renormalized polaron--polariton manifold
into $N_{\mathcal B}$ coupled-state subblocks.
The most demanding PB-VVPT situation corresponds to a single
coupled-state block, $N_{\mathcal B}=1$ with
$\dim(\mathcal B)=N+1$.
In this \textit{``worst-case scenario''},
Eq.~\eqref{eq:cost_effective_preparation} reduces to
\begin{equation}
C_{\text{\normalfont PB-VVPT}}^{\rm \normalfont\ worst}
 =
 \mathcal O\left[
 (N+1)^3
 \sum_{n=0}^{N_{\mathrm{ph}}^{\max}}(D^{(n)})  ^3
 \right].
 \label{eq:cost_effective_single_block}
\end{equation}
Comparing Eqs.~\eqref{eq:cost_exact_preparation}
and~\eqref{eq:cost_effective_single_block} shows that,
even in the single-block treatment, the decomposition
into fixed-phonon-number sectors provides a systematic
reduction in the estimated diagonalization cost (sum of cubic terms instead of a cubic sum).

\begin{figure}[t]
    \centering 
    \includegraphics[width=\columnwidth]{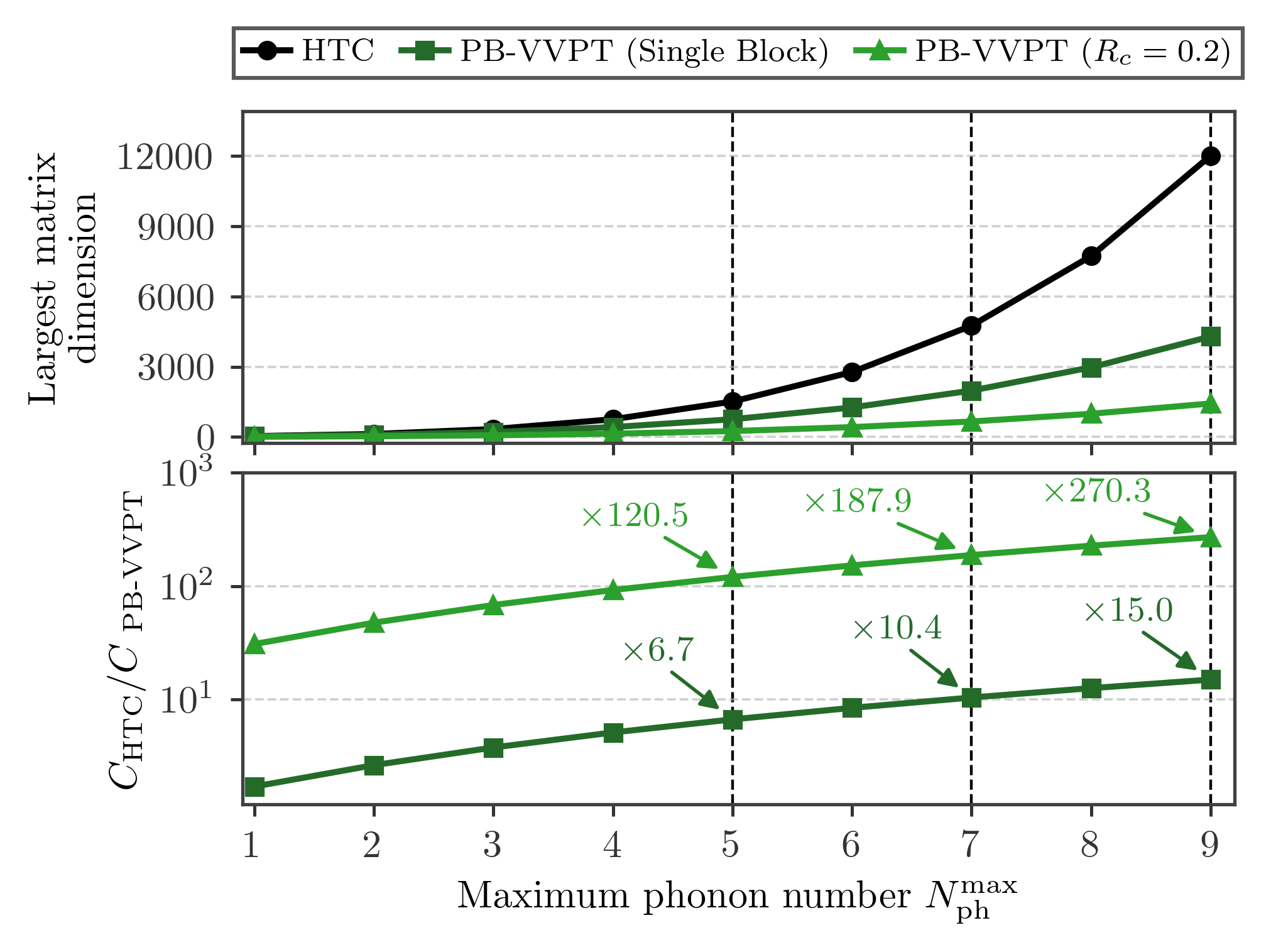}
\caption{
   Computational cost reduction achieved by PB-VVPT for the five-molecule wheel network as a function of the maximum total phonon number $N_{\mathrm{ph}}^{\max}$.
The upper panel compares the dimension of the full HTC matrix (black) with the largest matrix dimension among the independent PB-VVPT eigenvalue problems, for the single-block treatment (dark green) and the filtered partition with $R_{\mathrm c}=0.2$ (light green).
The lower panel shows the corresponding ratios $C_{\mathrm{HTC}}/C_{\mathrm{PB\text{-}VVPT}}$ of estimated total diagonalization costs, obtained by summing the cubic costs of all required dense-matrix diagonalizations in each approach.
The vertical dotted lines indicate the phonon cutoffs used in the numerical benchmarks in Sec.~\ref{sec:numerical_results}, namely $N_{\mathrm{ph}}^{\max}=5$, $7$, and $9$.}
    \label{fig:computational_cost}
\end{figure}

To better illustrate this advantage, Fig.~\ref{fig:computational_cost}
compares the largest matrix dimensions and the estimated
diagonalization costs for the exact HTC and PB-VVPT
approaches, considering a concrete example as given by the five-molecule network ($N=5$)  investigated
in Sec.~\ref{sec:numerical_results}.
Here, the filtered treatment with $R_c=0.2$ involves one
two-dimensional block and four one-dimensional blocks,
whose physical origin is discussed in the next section.
As shown in the upper panel, at the largest phonon cutoff considered,
$N_{\mathrm{ph}}^{\max}=9$, the HTC Hamiltonian involves
a matrix of dimension $12012$, compared with $4290$
for the largest matrix in the single-block PB-VVPT treatment
and only $1430$ for the filtered partition.
In the lower panel, we assess the reductions in diagonalization cost estimated
from the ratio of the cubic-scaling expressions in
Eqs.~\eqref{eq:cost_exact_preparation}
and~\eqref{eq:cost_effective_preparation}.
For the phonon cutoffs used in our numerical benchmarks,
these estimates yield reduction factors ranging from
 $6.7$ to $15$ in the single-block treatment (i.e. \textit{worst case scenario}),
while reaching $120.5$ to $270.3$ with  filtered partition.
These results highlight the substantial computational
advantage of PB-VVPT, with estimated reductions exceeding
two orders of magnitude through the combined decomposition
into fixed-phonon-number sectors and coupled-state subblocks.

To conclude, it is worth mentioning that the computational cost
of the PB-VVPT method can be reduced even further when all
renormalized polaron--polariton states are effectively decoupled.
In this limit, each state defines an independent block, with
$N_{\mathcal B}=N+1$ and $\dim(\mathcal B)=1$, and the associated
conditional phonon dynamics admits an analytical treatment
in terms of quadratic bosonic Hamiltonians specified by
$N\times N$ matrices.
The explicit fixed-$n$ many-body diagonalizations are therefore
no longer required, allowing the propagators to be evaluated
without constructing the truncated many-body phonon basis.
The corresponding analytical expressions and a brief discussion
of their computational cost are provided in
Appendix~\ref{app:nondegenerate_reduction}.

\begin{figure*}[t]
    \centering
    \includegraphics[width=16.5cm]{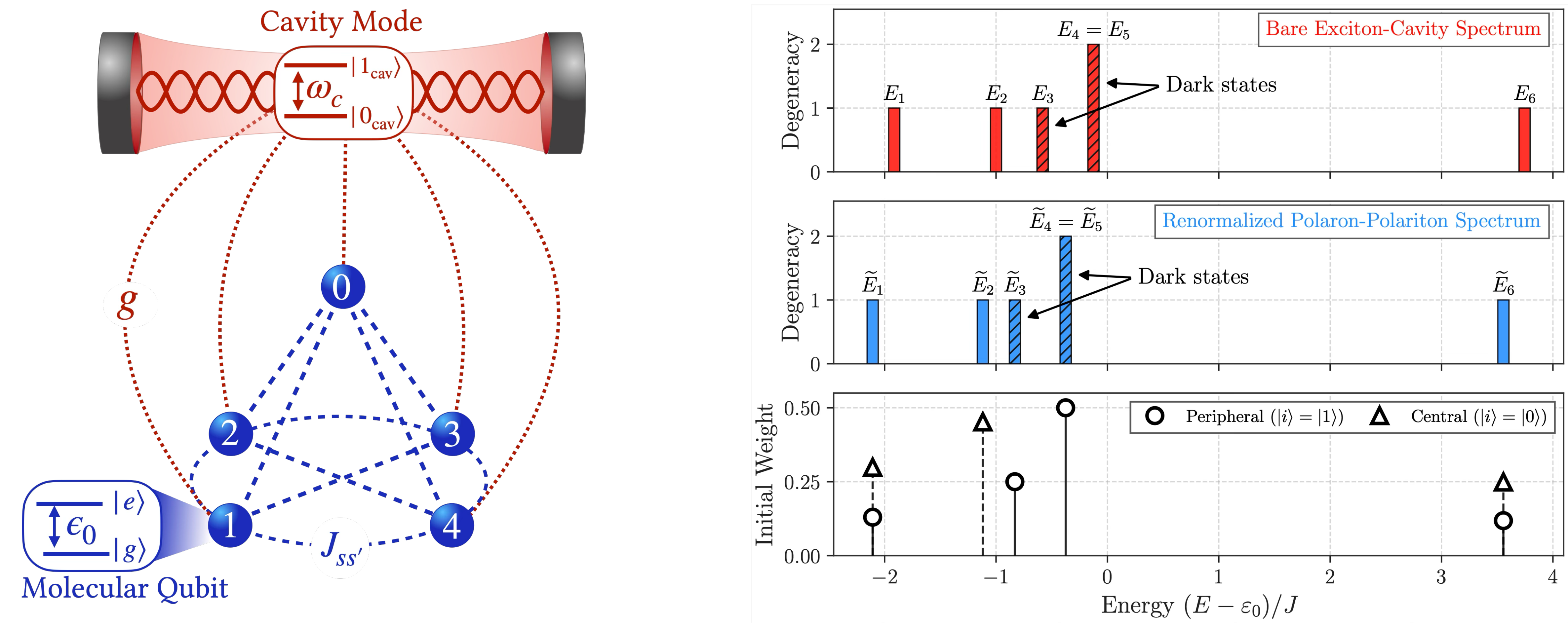}
\caption{
Illustration of the properties of the molecular wheel network used in our numerical benchmark of PB-VVPT.
Left: Schematic representation of the molecular network, consisting of one central site surrounded by four equally spaced peripheral sites. Excitonic interactions follow distance-dependent dipole--dipole couplings, with radial coupling strength $J$ and peripheral interactions $J_{ss'}=J/[8|\sin(\pi(s-s')/4)|^3]$. All molecular sites are uniformly coupled to a common cavity mode with strength $g$, while each molecule interacts with a local phonon mode (not shown).
Right: Discrete energy spectra of the bare exciton--cavity Hamiltonian (top) and the renormalized polaron--polariton Hamiltonian obtained within PB-VVPT (middle). Bar heights indicate the degeneracy of each level, while hatched bars identify dark excitonic states decoupled from the cavity by symmetry. The bottom panel shows the spectral weights $|\langle\widetilde{\chi}_\mu|i\rangle|^2$ of two initial molecular excitations localized on a peripheral site ($|i\rangle=|1\rangle$, circles) or the central site ($|i\rangle=|0\rangle$, triangles). Contributions from degenerate states are summed, and weights below $1\%$ are omitted for clarity.  Calculations are performed with $\Omega_0=100J$, $g=1J$, $\omega_c=\varepsilon_0$, and $\lambda=5J$. 
}
\label{fig:wheel_graph}
\end{figure*}

 \section{Numerical Benchmarks and Discussion}
\label{sec:numerical_results} 

We now benchmark the PB-VVPT method introduced above against
the exact dynamics generated by the full HTC Hamiltonian.
For this purpose, we consider a molecular network whose small size allows both Hamiltonians to be
diagonalized directly within the same truncated many-body space.
After introducing the network and its polaritonic spectrum,
we examine two different kinds of QST scenarios:
\textit{(i)} site-to-site QST between molecular qubits and
\textit{(ii)} site-to-cavity QST from a molecular qubit to
the cavity mode.
These configurations allow us to assess the accuracy of PB-VVPT
across different exciton--phonon coupling strengths and
temperatures, and to examine the role of the partition
into coupled-state blocks and light--matter coupling.
In the following, all numerical simulations are performed using the
\texttt{QuantNBody} package~\cite{yalouz2022quantnbody},
developed by one of the present authors, which provides
an exact matrix representation of the relevant operators
within the truncated many-body basis as discussed  previously in Section~\ref{sec:numerical_evaluation}.

 \subsection{Molecular Wheel Network and Polaritonic Spectra}

In our numerical benchmarks, we consider a molecular wheel network of $N=5$ sites, consisting of one central site surrounded by four equally spaced peripheral sites forming a closed ring, as illustrated in Fig.~\ref{fig:wheel_graph}.
This geometry provides a simple yet sufficiently rich setting
to examine both degenerate and nondegenerate states, as well
as transfer processes involving different contributions from
the bright and dark sectors.
Note that related architectures, with peripheral units surrounding
a central acceptor, have also been studied in optical cavities
in the context of excitation-energy
transfer~\cite{saez2019cavity,alexander2026photon}.

As illustrated in Fig.~\ref{fig:wheel_graph}, the molecular sites of the network are labeled $s=0,\ldots,4$, with $s=0$
denoting the central site.
All molecular sites have the same transition energy
$\varepsilon_0$ and couple uniformly to a common cavity
mode with strength $g$.
We consider distance-dependent dipole--dipole interactions
between molecular sites, taking the center-to-periphery
distance as the reference. All radial couplings are thus
set to $J$, while the couplings between distinct peripheral
sites $s$ and $s'$ are given by
\begin{equation}
J_{ss'}=\frac{J}{8|\sin[\pi(s-s')/4]|^3}.
\end{equation}
In the following, we use $J$ as the energy unit and consider
a cavity resonant with the molecular transition,
$\omega_c=\varepsilon_0$.
We fix $\Omega_0=100J$ and, unless otherwise stated,
take $g=J$ and $\lambda=5J$.
These parameters satisfy the perturbative conditions
of Eq.~\eqref{eq:perturbative_regime}, with
$\lambda/\Omega_0=0.05$ and
$\Delta E_{\mathrm{exc-cav}}/\Omega_0\sim0.06$.
 
The upper- and middle-right panels of Fig.~\ref{fig:wheel_graph} show the bare ($\lambda=0J$) and renormalized ($\lambda=5J$) polaritonic spectra, respectively.
Owing to the symmetry of the wheel, three states ($E_1$, $E_2$, and $E_6$) are bright, involving the central site, the symmetric peripheral excitation, and the cavity, while the remaining three are dark and localized within the peripheral subspace.
The dark sector comprises one nondegenerate state and an exactly degenerate doublet ($E_4=E_5$).
Phonon dressing induces state-dependent energy shifts while preserving this degeneracy ($\widetilde E_4=\widetilde E_5$).
For $R_{\mathrm c}=0.2$, the coupled-state partition introduced in Sec.~\ref{sec:effective_hamiltonian_construction} therefore consists of four one-dimensional blocks and one two-dimensional block associated with the dark doublet.
This partition is used throughout the following benchmarks
and compared with a single-block treatment retaining
all six polaron--polariton states.
The computational resource reductions provided by both partitions compared to the exact HTC treatment are quantified
in Fig.~\ref{fig:computational_cost}.


Finally, the bottom-right panel of Fig.~\ref{fig:wheel_graph}
shows the spectral weights $|\langle\widetilde\chi_\mu|i\rangle|^2$
of the initial states considered in the two QST scenarios below:
\textit{(i)} an excitation on a peripheral site,
$\ket{i}=\ket{1}$, and \textit{(ii)} an excitation on the central
site, $\ket{i}=\ket{0}$.
These weights quantify the contribution of each renormalized
polaron--polariton state to the initial excitation.
For site-to-site QST, the initial peripheral excitation carries
$75\%$ of its spectral weight in the dark sector.
By contrast, for site-to-cavity QST, the initial central-site
excitation lies entirely within the bright sector.
These two protocols therefore allow us to examine how
different contributions from the bright and dark sectors
shape QST in the presence of phonons.

\subsection{Site-to-Site Quantum-State Transfer}

As a first case study,
we consider QST between two diametrically opposite peripheral
sites of the wheel network, choosing $\ket{i}=\ket{1}$
and $\ket{f}=\ket{3}$ as the initial and target states,
respectively.
In this scenario, the initial state carries most of its weight in the dark
sector, with the degenerate doublet accounting for the largest
share (see Fig.~\ref{fig:wheel_graph}).
This configuration therefore allows us to test PB-VVPT
when degenerate states contribute substantially to
the transfer dynamics.
We first examine the short- and long-time dynamics at
$\lambda=5J$ using the block partition introduced above,
highlighting the emergence of thermal decoherence and
comparing the PB-VVPT predictions with the exact HTC results.
We then assess the agreement quantitatively by examining
how the errors in the transfer fidelity depend on
temperature, exciton--phonon coupling, and block filtering.
\subsubsection{Transfer Dynamics and Thermal Decoherence}

\begin{figure}[t]
    \centering
    \includegraphics[width=7.5cm]{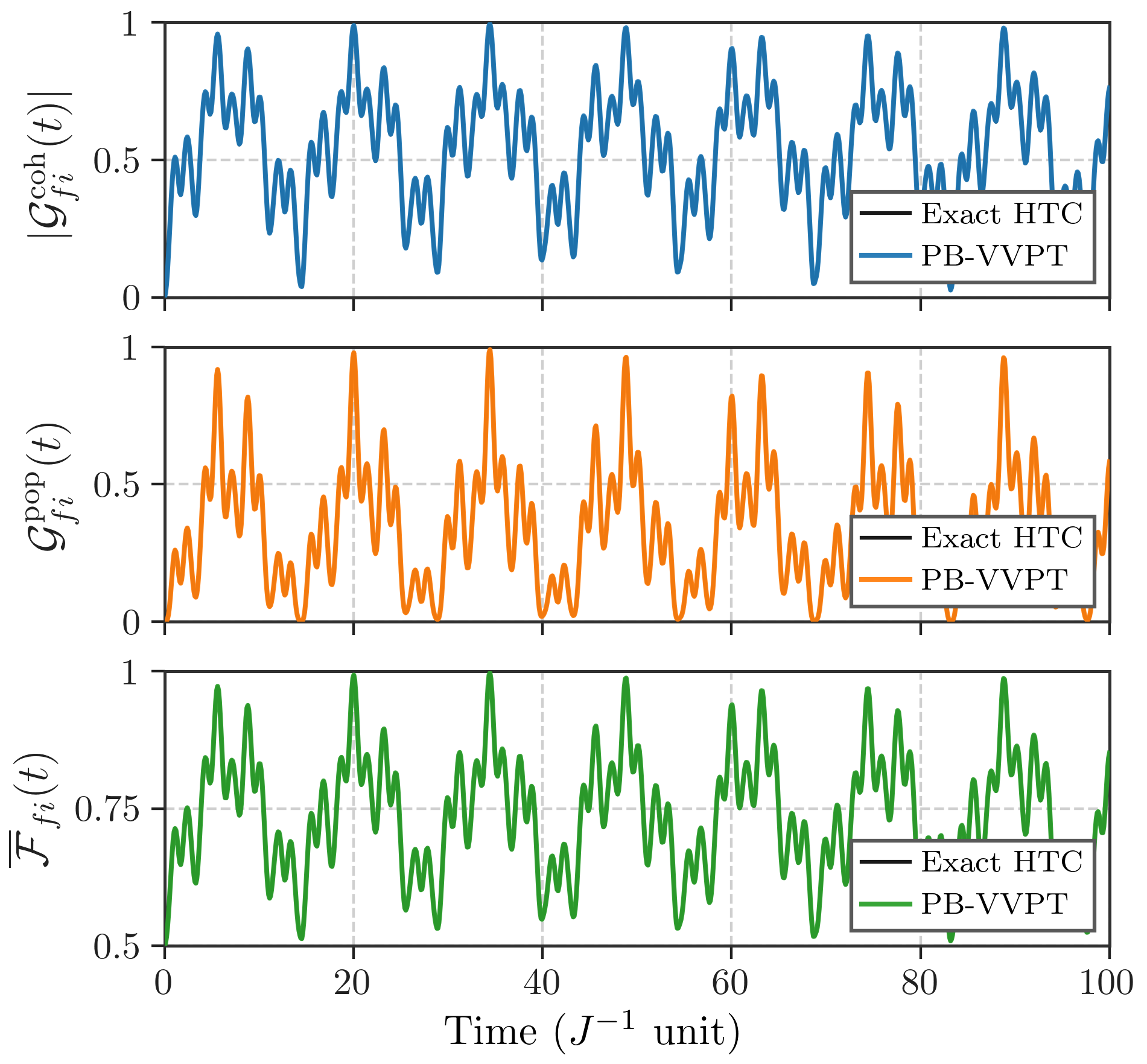}
    \caption{
    Short-time quantum-state-transfer dynamics between
    diametrically opposite peripheral sites of the molecular
    wheel network, $\ket{i}=\ket{1}$ and $\ket{f}=\ket{3}$.
    From top to bottom, the panels show the coherence
    propagator $|\mathcal{G}_{fi}^{\mathrm{coh}}(t)|$,
    the population propagator
    $\mathcal{G}_{fi}^{\mathrm{pop}}(t)$, and the average
    transfer fidelity $\overline{\mathcal{F}}_{fi}(t)$.
    Black solid curves represent exact HTC dynamics,
    while colored solid curves correspond to filtered
    PB-VVPT predictions with $R_c=0.2$.
    Calculations are realized here with $\lambda=5J$, $\Omega_0=100J$,
    $g=1J$, $\omega_c=\varepsilon_0$, and
    $k_{\mathrm B}T/\Omega_0=0.25$, with a maximum
    total phonon occupation $N_{\mathrm{ph}}^{\max}=5$.
    }
    \label{fig:qst_short}
\end{figure}

In Fig.~\ref{fig:qst_short}, we first illustrate the transfer
dynamics over the time interval $t\in[0,100J^{-1}]$
through the time evolution of the coherence propagator
$|\mathcal{G}_{fi}^{\mathrm{coh}}(t)|$,
the population propagator
$\mathcal{G}_{fi}^{\mathrm{pop}}(t)$, and the
transfer fidelity $\overline{\mathcal{F}}_{fi}(t)$
defined in Eq.~\eqref{eq:average_transfer_fidelity_main}.
The temperature is set to $k_{\mathrm B}T/\Omega_0=0.25$,
with a maximum total phonon number
$N_{\mathrm{ph}}^{\max}=5$.
As readily seen in the figure, throughout this time interval
the PB-VVPT predictions (colored curves) closely overlap
the exact HTC results (black curves), reproducing the
oscillatory dynamics of all three quantities characterizing
the QST process.
Interestingly, we observe that high-fidelity transfer occurs
repeatedly, with several maxima around $t\simeq20J^{-1}$,
$35J^{-1}$, and $50J^{-1}$, where both propagators approach unity.
Additional simulations (not shown here) reveal that varying
the light--matter coupling strength $g$ only slightly affects
the overall transfer dynamics, mainly modulating the timing
and amplitude of these fidelity maxima.
This relatively weak dependence on $g$ is consistent with
the predominant spectral weight of the initial state in the
dark sector, as shown previously (see Fig.~\ref{fig:wheel_graph}).
By contrast, as we will show in Sec.~\ref{sec:QST_light_matter},
the influence of $g$ becomes more pronounced for
site-to-cavity transfer, where the initial state lies entirely
in the bright sector.

\begin{figure*}[t]
    \centering
    \includegraphics[width=16cm]{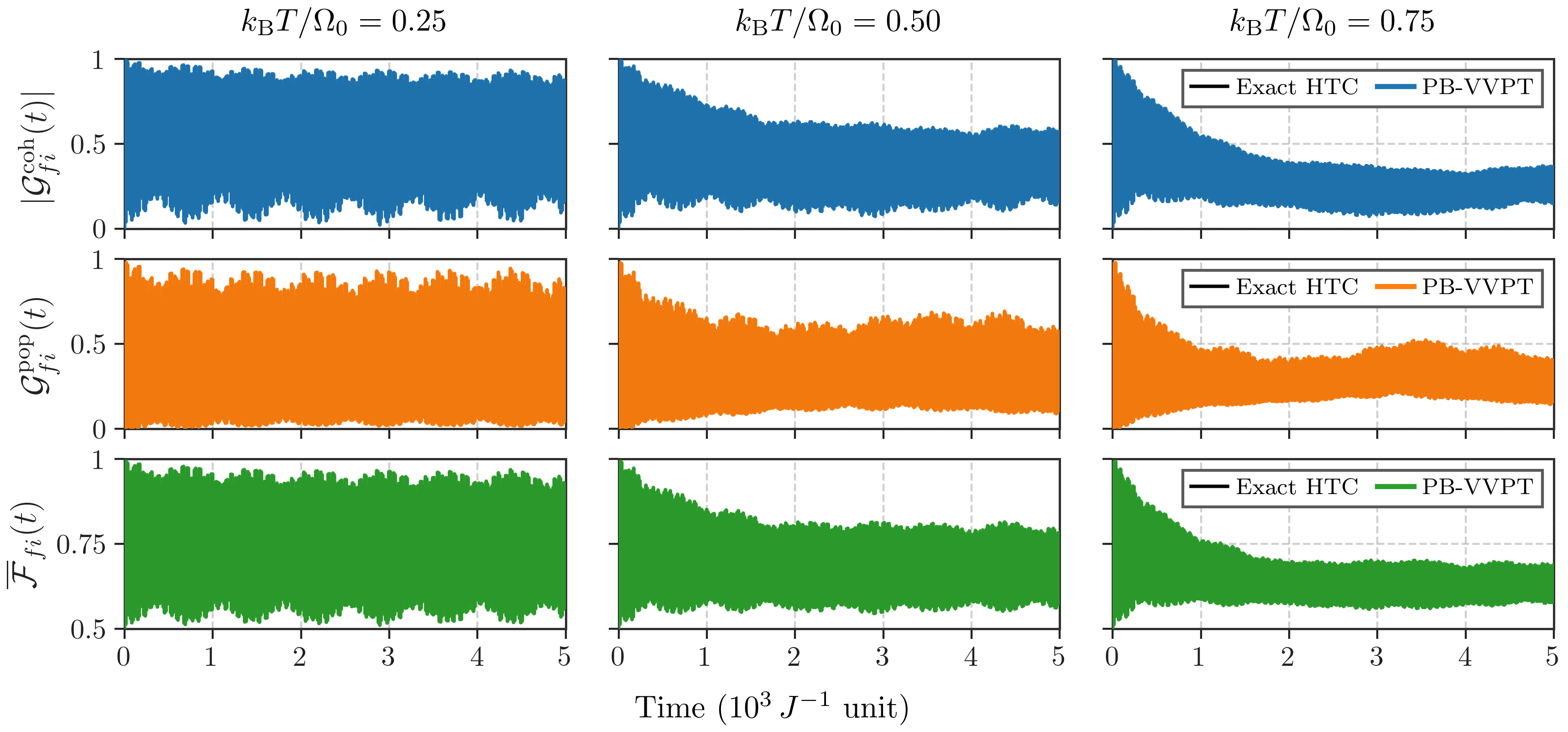}
    \caption{
    Temperature dependence of long-time quantum-state
    transfer between the peripheral sites
    $\ket{i}=\ket{1}$ and $\ket{f}=\ket{3}$.
    From top to bottom, the rows show the coherence
    propagator $|\mathcal{G}_{fi}^{\mathrm{coh}}(t)|$,
    the population propagator
    $\mathcal{G}_{fi}^{\mathrm{pop}}(t)$, and the average
    transfer fidelity $\overline{\mathcal{F}}_{fi}(t)$.
    Columns correspond to
    $k_{\mathrm B}T/\Omega_0=0.25$, $0.50$, and $0.75$,
    from left to right, with respective total phonon
    cutoffs $N_{\mathrm{ph}}^{\max}=5$, $7$, and $9$.
    Black solid curves represent exact HTC dynamics,
    while colored solid curves correspond to filtered
    PB-VVPT predictions with $R_c=0.2$.
    Parameters are $\lambda=5J$, $\Omega_0=100J$,
    $g=J$, and $\omega_c=\varepsilon_0$.}
    \label{fig:qst_long}
\end{figure*}

Turning now to the effects of temperature, thermal decoherence
remains barely visible over this short time interval.
To examine how it develops and affects the transfer,
we extend the simulations to $t=5000J^{-1}$ and compare
three temperatures $k_{\mathrm B}T/\Omega_0=0.25$, $0.50$, and $0.75$, in Fig.~\ref{fig:qst_long}.
The corresponding total phonon cutoffs are $N_{\mathrm{ph}}^{\max}=5$, $7$, and $9$, respectively, with the same cutoff used in the HTC and PB-VVPT calculations at each temperature.
These cutoffs keep the omitted thermal weight $1-Z_{\mathrm{ph}}^{\mathrm{trunc}}/Z_{\mathrm{ph}}$ below $0.1\%$ in all three cases.
For an illustrative analysis of the convergence of the transfer dynamics with respect to the phonon cutoff, we refer the reader to Appendix~\ref{app:convergence}.

At the lowest temperature, $k_{\mathrm B}T/\Omega_0=0.25$
(left column of Fig.~\ref{fig:qst_long}), all three quantities
retain large oscillations throughout the simulation.
The population and coherence propagators repeatedly recover
large values, allowing high-fidelity transfer even at long times.
However, as the temperature increases, these oscillations become
progressively attenuated.
For $k_{\mathrm B}T/\Omega_0=0.50$ (middle column),
this attenuation develops over the first few thousand
units of $J^{-1}$, reducing the transfer fidelity.
At $k_{\mathrm B}T/\Omega_0=0.75$ (right column),
the attenuation is more pronounced, with all three quantities
subsequently oscillating within narrower ranges and the
fidelity reaching lower maxima.
These long-time dynamics reveal the effects of phonon-induced
decoherence, which become more pronounced as temperature
increases and limit the recovery of quantum information
at the target site.

Importantly, as shown in Fig.~\ref{fig:qst_long},
all PB-VVPT results closely follow the exact HTC dynamics
at all three temperatures, with curves that are nearly
indistinguishable over the entire time interval considered.
This excellent agreement shows that PB-VVPT accurately
reproduces both the thermal attenuation and the high-frequency
oscillations persisting at very long times.
However, such a visual comparison does not quantify the
residual deviations from the exact HTC dynamics.
In the following subsection, we therefore examine these
numerical deviations more quantitatively by evaluating the errors
in the average transfer fidelity and assessing the influence
of the block treatment introduced in
Sec.~\ref{sec:effective_hamiltonian_construction}.

\subsubsection{ Accuracy of PB-VVPT and Block-Filtering }

\begin{figure*}[t]
    \centering
    \includegraphics[width=\textwidth]{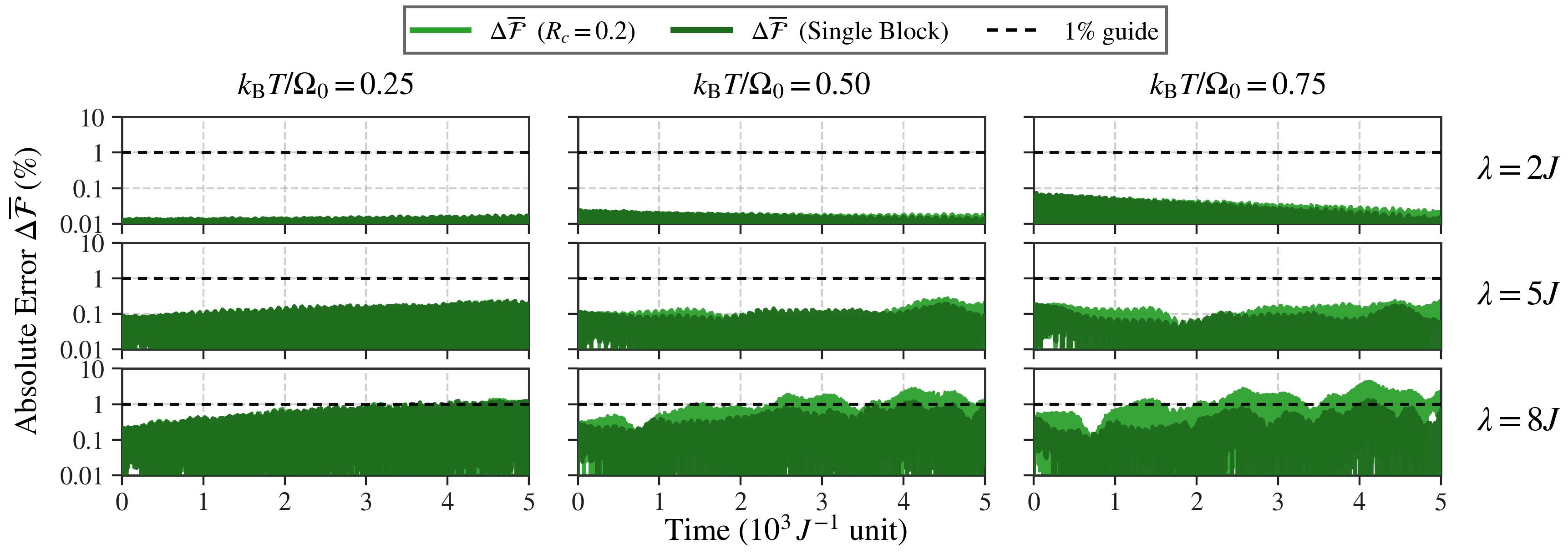}
    \caption{
    Absolute errors in the average qubit-transfer fidelity
    relative to the exact HTC dynamics, for transfer between
    opposite peripheral sites $\ket{i}=\ket{1}$ and
    $\ket{f}=\ket{3}$.
    Light green curves correspond to PB-VVPT with the
    projected-block approximation using $R_c=0.2$,
    while dark green curves correspond to the single-block
    treatment retaining all six renormalized
    polaron--polariton states.
    Rows correspond to $\lambda/J=2$, $5$, and $8$
    from top to bottom, and columns to
    $k_{\mathrm B}T/\Omega_0=0.25$, $0.50$, and $0.75$
    from left to right, with respective total phonon
    cutoffs $N_{\mathrm{ph}}^{\max}=5$, $7$, and $9$.
    The dashed black line marks
    $\Delta\overline{\mathcal F}_{fi}=1\%$.
    The remaining parameters are $\Omega_0=100J$,
    $g=J$, and $\omega_c=\varepsilon_0$.
    }
    \label{fig:qst_fidelity_errors}
\end{figure*}

To quantify the accuracy of PB-VVPT against the exact HTC
dynamics, we consider the absolute error in the average
transfer fidelity, defined as (in percentage):
\begin{equation}
 \Delta\overline{\mathcal F}_{fi}(t)
 =
 100\left|
 \overline{\mathcal F}_{fi}^{\mathrm{PB\text{-}VVPT}}(t)
 -
 \overline{\mathcal F}_{fi}^{\mathrm{HTC}}(t)
 \right|,
 \label{eq:error_fidelity}
\end{equation}
where here PB-VVPT denotes either the filtered or single-block
treatment which we will compare in the following.
For all parameters considered here, the filtering threshold
$R_c=0.2$ yields the same partition: four one-dimensional
blocks and one two-dimensional block containing the
degenerate dark doublet.
In contrast, the \textit{single-block} treatment retains all six
renormalized polaron--polariton states and their residual
phonon-number-conserving interactions, allowing us to assess
the influence of the neglected interblock couplings
(see Fig.~\ref{fig:computational_cost} for a comparison of
their computational costs).

Figure~\ref{fig:qst_fidelity_errors} shows the resulting
fidelity errors over $t\in[0,5000J^{-1}]$ for increasing
exciton--phonon coupling (rows) and temperature (columns).
At weak and intermediate couplings, $\lambda=2J$ and $5J$,
both PB-VVPT treatments remain highly accurate across
all temperatures, with errors below $0.1\%$ and limited
to a few tenths of a percent, respectively.
The two treatments nearly coincide at $\lambda=2J$,
indicating that interblock couplings have little influence
in this regime.
At $\lambda=5J$, their differences become more visible
as the temperature increases, although the filtered treatment
continues to reproduce the exact HTC dynamics with
excellent accuracy.

At stronger coupling, $\lambda=8J$, the influence of the projected-block approximation becomes more pronounced, particularly at long times and higher temperatures. The filtered-treatment error reaches the $1\%$ threshold at approximately $t=3000J^{-1}$, $1500J^{-1}$, and $1000J^{-1}$ for $k_{\mathrm B}T/\Omega_0=0.25$, $0.50$, and $0.75$, respectively. At the highest temperature, subsequent deviations reach a few percent, whereas the single-block treatment maintains errors around or below $1\%$ over most of the interval. This improvement highlights the increasing contribution of neglected interblock couplings in this regime.

Overall, these benchmarks demonstrate that PB-VVPT accurately reproduces QST dynamics over timescales far exceeding $J^{-1}$. The projected-block approximation provides a substantial reduction in numerical cost while retaining sub-percent fidelity errors over broad parameter and time ranges. At stronger exciton--phonon coupling and higher temperatures, retaining the full set of residual interblock couplings becomes beneficial for preserving long-time accuracy.

\begin{figure*}[t]
    \centering
    \includegraphics[width=\textwidth]{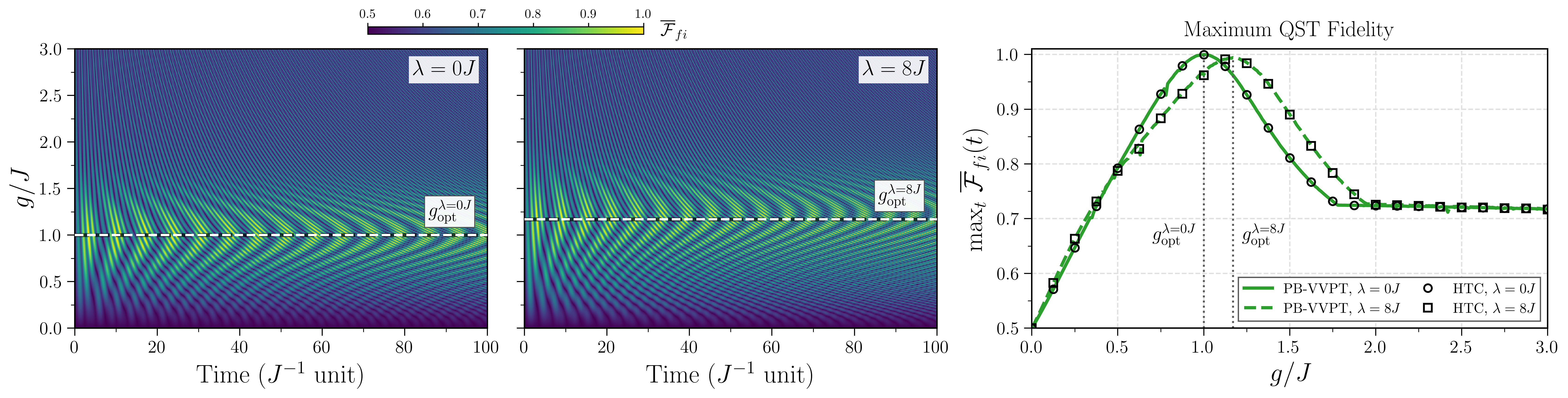}
    \caption{
    Short-time matter-to-light quantum-state transfer from the
    central site of the molecular wheel, $\ket{i}=\ket{0}$,
    to the cavity mode, $\ket{f}=\ket{c}$.
    The left and middle panels show the average transfer fidelity
    $\overline{\mathcal F}_{fi}(t)$ predicted by PB-VVPT
    as a function of time and light--matter coupling $g/J$,
    for $\lambda/J=0$ and $8$, respectively.
    White horizontal dashed lines indicate the optimal couplings
    $g_{\mathrm{opt}}$, which maximize the fidelity over
    the interval $t\in[0,100J^{-1}]$.
    The right panel shows the maximum fidelity over the same
    interval as a function of $g/J$.
    Green solid and dashed curves correspond to PB-VVPT
    for $\lambda/J=0$ and $8$, respectively, while black
    open circles and squares show the corresponding exact HTC results.
    Vertical dotted lines mark the optimal couplings
    $g_{\mathrm{opt}}^{0J}$ and $g_{\mathrm{opt}}^{8J}$,
    where the superscript specifies the exciton--phonon
    coupling strength.
    The remaining parameters are $\Omega_0=100J$,
    $\omega_c=\varepsilon_0$, and $k_{\mathrm B}T/\Omega_0=0.25$,
    with a total phonon cutoff $N_{\mathrm{ph}}^{\max}=5$
    and a single-block treatment. 
    }
    \label{fig:Perfect_QST_g}
\end{figure*}

\subsection{Site-to-Cavity Quantum-State Transfer}
\label{sec:QST_light_matter}

Having investigated matter-to-matter QST between peripheral molecular sites, we now turn to matter-to-light transfer from the central molecular site, $|i\rangle=|0\rangle$, to the cavity mode, $|f\rangle=|c\rangle$.
Similar matter--light quantum-state transfer protocols have also been investigated in other contexts (see Refs.~\cite{rabl2006hybrid,thijssen2012transfer,bechler2018passive}).
In our case, as illustrated in Fig.~\ref{fig:wheel_graph}, the initial state has its entire spectral weight in the nondegenerate bright sector, in contrast to the previous configuration where dark states contribute substantially to the dynamics.


\subsubsection{Short-Time Transfer: Optimal Light--Matter Coupling and Phonon-Induced Renormalization}

We first investigate how the light--matter coupling strength $g$ controls short-time QST and how the transfer conditions are modified by exciton--phonon coupling. Figure~\ref{fig:Perfect_QST_g} presents the average transfer fidelity over $t\in[0,100J^{-1}]$ for $\lambda=0$ and $\lambda=8J$. Both calculations are performed at $k_{\mathrm B}T/\Omega_0=0.25$, using a total phonon cutoff $N_{\mathrm{ph}}^{\max}=5$ with a single-block treatment of the renormalized polaron--polariton subspace.

The PB-VVPT fidelity heatmaps in the left and middle panels reveal a region of optimal light--matter coupling around $g\simeq J$, characterized by repeated high-fidelity transfer events (bright yellow regions). Exciton--phonon coupling modifies the temporal pattern of these events, reflecting changes in the frequencies governing the transfer dynamics. To identify the optimal coupling in each case, we evaluate the maximum fidelity over the time interval considered for each value of $g$. The resulting optimal couplings are marked by white horizontal dashed lines in the heatmaps and vertical dotted lines in the rightmost panel.

Focusing on the rightmost panel, the maximum fidelities obtained with PB-VVPT (green curves) are compared with the exact HTC results (black open markers), showing excellent agreement over the full range of light--matter couplings considered. In the absence of exciton--phonon coupling (solid green curve and black open circles), the maximum fidelity approaches unity at $g_{\mathrm{opt}}^{\lambda=0}=J$. Actually, this optimal condition admits an analytical interpretation based on the symmetry of the exciton--cavity Hamiltonian, which we detail for completeness in Appendix~\ref{app:Anlaytical_optimal_QST}. Increasing the light--matter coupling beyond this optimum does not improve the transfer. Instead, the maximum fidelity decreases and approaches values around $0.72$ for $g\gtrsim2J$.

For $\lambda=8J$ (dashed green curve and black open squares), the maximum shifts to $g_{\mathrm{opt}}^{\lambda=8J}\simeq1.2J$, with a fidelity close to $0.99$. Thus, although exciton--phonon coupling reduces the transfer fidelity at $g=J$, we clearly see here that near-unit fidelity can be recovered by adjusting the light--matter coupling.

\begin{figure*}[t]
    \centering
    \includegraphics[width=16.5cm]{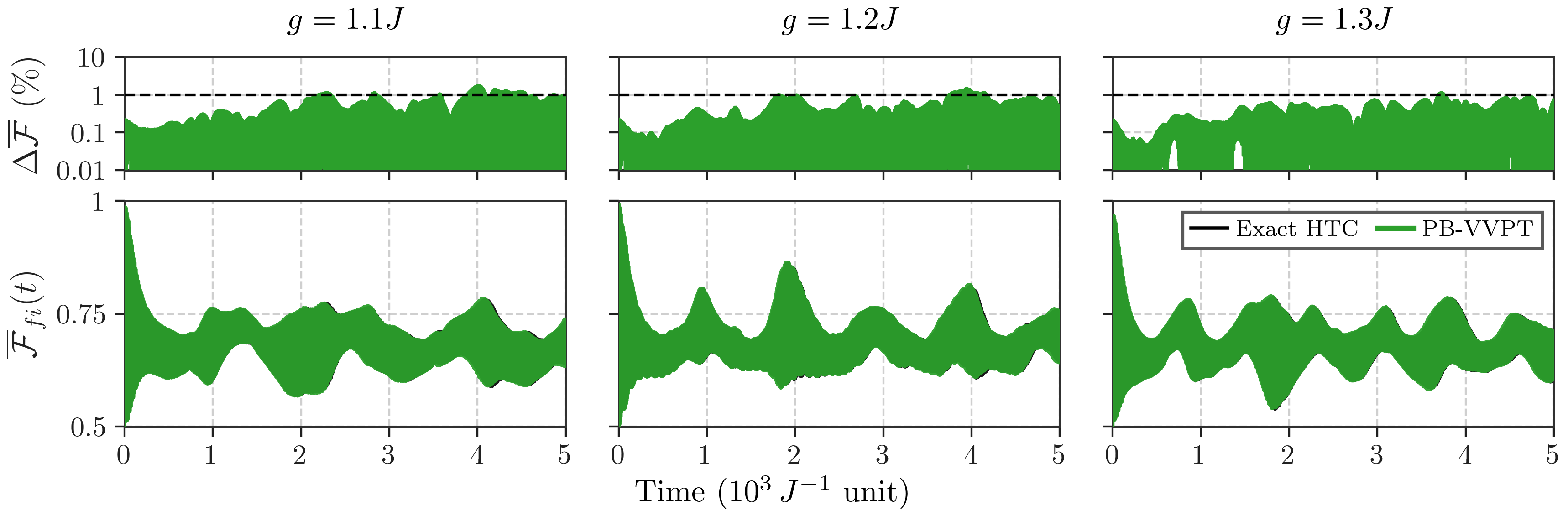}
\caption{
Long-time dynamics of the QST from the central molecular site  to the cavity mode, for $g/J=1.1$ (left), $1.2$ (middle), and $1.3$ (right).
The exciton--phonon coupling and temperature are fixed at $\lambda=8J$ and $k_{\mathrm B}T/\Omega_0=0.75$, respectively, with a total phonon cutoff $N_{\mathrm{ph}}^{\max}=9$.
PB-VVPT calculations are realized with a single block retaining all renormalized polaron--polariton states.
The upper panels show the absolute error in the average transfer fidelity $\Delta \overline{\mathcal{F}} (t)$ .  
The lower panels show the average qubit-transfer fidelity $\overline{\mathcal{F}}_{fi}(t)$, with exact HTC results in black and PB-VVPT predictions in green.
The long-time revivals exhibit coupling-dependent amplitudes and recurrence times.
}
\label{fig:long_time_transduction} 
\end{figure*}

\subsubsection{Long-Time Transfer: \linebreak Fidelity Revivals and Non-Markovian Decoherence }

We now extend the analysis to long times to examine how the light--matter coupling affects QST in the presence of thermal decoherence. Figure~\ref{fig:long_time_transduction} shows the average transfer fidelity over $t\in[0,5000J^{-1}]$ for $g/J=1.1$, $1.2$, and $1.3$, at fixed $\lambda=8J$ and $k_{\mathrm B}T/\Omega_0=0.75$. These couplings are chosen around the short-time optimum identified above (see Fig.~\ref{fig:Perfect_QST_g}). The lower panels compare the single-block PB-VVPT results with the exact HTC dynamics, while the upper panels show the corresponding fidelity errors $\Delta\overline{\mathcal{F}}$ as defined in Eq.~\eqref{eq:error_fidelity}.


As shown in the lower panels of Fig.~\ref{fig:long_time_transduction},
thermal decoherence initially attenuates the transfer oscillations
over a characteristic timescale of approximately $500J^{-1}$.
At longer times, however, pronounced fidelity revivals emerge
despite the relatively high temperature.
Similar behavior was extensively investigated in our previous
studies of QST in excitonic networks without
cavities~\cite{yalouz2017exciton,yalouzQInf17b}, where partial
recurrences of the phonon dynamics were identified as the origin
of non-Markovian decoherence and coherence revivals.
In light of these earlier findings, the revivals observed here
can naturally be attributed to the same phonon-induced
non-Markovian effects, but now within the polaritonic framework.

Interestingly, both the amplitude and the recurrence times of these revivals depend strongly on the light--matter coupling $g$. For $g=1.1J$, the fidelity exhibits broad revivals of moderate amplitude, whereas at $g=1.2J$ a more pronounced recovery occurs near $t\simeq2000J^{-1}$, reaching approximately $0.85$, followed by another revival near $t\simeq4000J^{-1}$. 
Increasing the coupling to $g=1.3J$ instead produces smaller and more fragmented revivals. The strongest long-time recovery therefore occurs near the coupling that also optimizes short-time QST (see Fig.~\ref{fig:Perfect_QST_g}), showing that cavity coupling can influence both the initial transfer efficiency and the subsequent non-Markovian revivals at longer times.

To conclude here, we note once again that PB-VVPT closely reproduces the exact HTC dynamics throughout these simulations, as confirmed by the fidelity errors reported in the upper panels of Fig.~\ref{fig:long_time_transduction}, which remain mostly below $1\%$ over the full time interval. This agreement demonstrates the ability of PB-VVPT to capture long-time fidelity revivals consistent with non-Markovian memory effects, including their dependence on the light--matter coupling strength.

\section{Conclusion }
\label{sec:conclusion}

In this work, we have developed an effective-Hamiltonian approach to investigate quantum-state transfer in molecular networks coupled to an optical cavity and local phonons. Starting from the generalized Holstein--Tavis--Cummings Hamiltonian, we have introduced a second-order effective description based on quasidegenerate Van Vleck perturbation theory, referred to as PB-VVPT. In the nonadiabatic regime of weak exciton--phonon coupling, this approach provides access to the coherence and population propagators governing quantum-state transfer, without invoking a Markovian approximation. By decomposing the dynamics into independent fixed-phonon-number sectors and polaron--polariton subblocks, PB-VVPT substantially reduces the computational cost while retaining the phonon-induced renormalization of the polaritonic states and their dynamics.

We have assessed the accuracy of PB-VVPT against exact reference calculations for a molecular wheel network, considering both site-to-site and site-to-cavity quantum-state transfer. These comparisons show that PB-VVPT accurately describes the influence of temperature, exciton--phonon coupling, and light--matter coupling over times extending far beyond the characteristic excitonic timescale. In particular, it reproduces the attenuation of transfer oscillations through thermal decoherence and the fidelity revivals observed at longer times. For site-to-cavity transfer, we further show that near-unit fidelity can be recovered by tuning the light--matter coupling, despite the presence of exciton--phonon interactions. Importantly, the benchmarks clarify the role of the projected-block approximation. Neglecting interblock couplings preserves accurate predictions over a broad time window, while retaining them improves the long-time agreement at stronger exciton--phonon coupling and higher temperature. Even in this single-block treatment, the decomposition into fixed-phonon-number sectors reduces the estimated diagonalization cost by approximately one order of magnitude for the phonon cutoffs considered, with reductions exceeding two orders of magnitude when block filtering is applied.


Naturally, the present work is a first step in the development and application of the PB-VVPT method. Having established its accuracy against exact reference calculations, the next objective is to apply it to situations that are beyond the reach of direct diagonalization of the full HTC Hamiltonian. This work also motivates a series of questions that could represent interesting starting points for future work. A first direction would be to extend the description beyond the rotating-wave approximation, for instance by considering a Pauli–Fierz Hamiltonian, to explore transfer processes in which the total excitation number is no longer conserved. Including higher photon-number sectors would also allow us to assess the validity of the two-level photonic encoding and investigate matter-to-light transfer as a qubit–qumode process, in connection with our recent work on polaritonic encodings~\cite{chiari2025ab}. A second direction would be to account for the finite cavity lifetime and examine how photon losses affect the transfer fidelity and its long-time revivals. Finally, applying the method to larger molecular networks with different geometries would allow us to investigate the influence of network architecture on quantum-information transfer. All of these developments are left for future work.

\begin{acknowledgments}
    This work benefited from State support managed by the ANR under the France 2030 program, referenced by ANR-23-PETQ-0006. 
\end{acknowledgments}

\appendix

\section{Derivation of the Van~Vleck Hamiltonian  }
\label{app:van_vleck_expansion}

In this Appendix, we detail the perturbative construction of the effective Van~Vleck Hamiltonian introduced in Eq.~\eqref{eq:effective_hamiltonian_second_order}, up to second order in the exciton--phonon coupling.
Starting from the partition defined in Eq.~\eqref{eq:perturbative_partition},
\begin{equation}
H=H_0+V 
\end{equation}
with $H_0=H_{\mathrm{exc-cav}}+H_{\mathrm{ph}}$ and
$V=H_{\mathrm{exc-ph}}$, 
we then introduce the unitary transformation
\begin{equation}
\widehat H=U_{\rm vv} H U_{\rm vv}^\dagger \quad \text{with} \quad U_{\rm vv} = e^S
\end{equation}
where the anti-Hermitian generator $S^\dagger=-S$ is expanded perturbatively as
\begin{equation}
S=S^{(1)}+S^{(2)}+\cdots,
\label{eq:appendix_generator_expansion}
\end{equation}
with $S^{(n)}$ denoting the $n$th-order contribution in the exciton--phonon coupling, i.e., $S^{(n)}=\mathcal O(\lambda^n)$. Using the Baker--Campbell--Hausdorff expansion
\begin{equation}
\widehat H
=
H+[S,H]
+\frac{1}{2}[S,[S,H]]
+\cdots,
\label{eq:appendix_bch}
\end{equation}
the transformed Hamiltonian can be organized order by order in the exciton--phonon coupling. Up to second order, we have
\begin{equation}
\begin{aligned}
\widehat H
&= H_0 \\
&+ V+[S^{(1)},H_0] \\
&+[S^{(2)},H_0]
+[S^{(1)},V]
+\frac{1}{2}[S^{(1)},[S^{(1)},H_0]].
\end{aligned}
\label{eq:appendix_bch_second_order}
\end{equation}
where each line of Eq.~\eqref{eq:appendix_bch_second_order} corresponds, respectively, to the zeroth-, first-, and second-order contributions.
At first order, the transformation is chosen to eliminate the exciton--phonon interaction, which changes the total phonon number by one. The corresponding condition is therefore
\begin{equation}
[H_0,S^{(1)}]=V.
\label{eq:appendix_first_order_condition}
\end{equation}
To satisfy this condition while explicitly preserving the anti-Hermiticity of the generator, we write
\begin{equation}
S^{(1)}
=
\sum_{s=1}^{N}
\left(
T_s b_s^\dagger
-
T_s^\dagger b_s
\right),
\label{eq:first_order_generator}
\end{equation}
where $T_s$ acts only on the exciton--cavity degrees of freedom. Using the spectral decomposition of $H_{\mathrm{exc-cav}}$ introduced in Eq.~\eqref{eq:polariton_eigenproblem}, it can be shown that Eq.~\eqref{eq:appendix_first_order_condition} is fulfilled for
\begin{equation}
T_s
=
\sum_{k,k'=1}^{N+1}
\frac{M_{kk'}^{(s)}}
{E_k-E_{k'}+\Omega_0}
\ketbra{\chi_k}{\chi_{k'}}.
\label{eq:one_phonon_transition_operator}
\end{equation}
Equivalently,
\begin{equation}
T_s^\dagger
=
\sum_{k,k'=1}^{N+1}
\frac{M_{kk'}^{(s)}}
{E_{k'}-E_{k}+\Omega_0}
\ketbra{\chi_k}{\chi_{k'}},
\end{equation}
where $M_{kk'}^{(s)}$ are the phonon-mediated polaritonic coupling matrix elements introduced in Eq.~\eqref{eq:polariton_phonon_matrix_elements}.
The energy denominators encode off-resonant virtual processes involving the creation or annihilation of one phonon.
Using Eq.~\eqref{eq:appendix_first_order_condition}, the second-order contribution in Eq.~\eqref{eq:appendix_bch_second_order} reduces to
\begin{equation}
\widehat H
=
H_0+[S^{(2)},H_0]
+\frac{1}{2}[S^{(1)},V].
\label{eq:appendix_second_order_reduced}
\end{equation}
The commutator $[S^{(1)},V]$ contains both phonon-number-conserving terms and contributions changing the total phonon number by two. In the Van Vleck construction, the latter are eliminated by the higher-order generator $S^{(2)}$. Its explicit form is not required here, since only the phonon-number-conserving contribution enters the effective Hamiltonian at the retained order. We can therefore write
\begin{equation}
\widehat H
=
H_0
+
\frac{1}{2}
\left[S^{(1)},V\right]_{\Delta n_{\mathrm{ph}}=0}.
\label{eq:appendix_effective_second_order}
\end{equation}

Normal ordering of the bosonic operators separates this remaining contribution into a phonon-independent correction to the exciton--cavity Hamiltonian and a quadratic phonon-number-conserving term, such as
\begin{equation}
\begin{aligned}
\frac{1}{2}
\left[S^{(1)},V\right]_{\Delta n_{\mathrm{ph}}=0}
={}&
\delta H_{\mathrm{exc-cav}}
\\
&+
\sum_{s,s'=1}^{N}
\sum_{k,k'=1}^{N+1}
\Lambda_{ss'}^{kk'}
\ketbra{\chi_k}{\chi_{k'}}
\otimes b_s^\dagger b_{s'} .
\end{aligned}
\label{eq:appendix_second_order_contribution}
\end{equation}
The matrix elements of $\delta H_{\mathrm{exc-cav}}$ and the coefficients $\Lambda_{ss'}^{kk'}$ are precisely those given in Eqs.~\eqref{eq:explicit_polariton_correction} and \eqref{eq:explicit_phonon_correction}, respectively. Putting these terms back into  Eq.~\eqref{eq:appendix_effective_second_order} yields the effective Hamiltonian of Eq.~\eqref{eq:effective_hamiltonian_second_order}.

The transformed Hamiltonian therefore conserves the total phonon number up to the retained perturbative order, although this quantity is not conserved by the original HTC Hamiltonian $H$. The resulting expressions constitute the direct exciton--cavity generalization of the second-order corrections previously obtained for exciton--phonon networks in Ref.~\cite{yalouz2017exciton}, with the cavity dependence entering through the polaritonic energies $E_k$ and the molecular amplitudes contained in $M_{kk'}^{(s)}$.

\section{Averaged Fidelity of Quantum-State Transfer  }
\label{app:fidelity}

In this appendix, we provide a short derivation of the average fidelity used to characterize the transfer of an arbitrary quantum state from $\ket{i}$ to $\ket{f}$. 
Starting from the pure qubit state $\ket{\psi_i}$ defined in Eq.~\eqref{eq:initial_qubit_state}, we define its logical counterpart by replacing $\ket{i}$ with $\ket{f}$:
\begin{equation}
\ket{\psi_{f\leftarrow i}}
=
c_0\ket{0_{\mathrm{exc-cav}}}+c_1\ket{f}.
\end{equation}
The similarity between this target state and the time-evolved reduced state $\rho_f(t)$ given in Eq.~\eqref{eq:final_qubit_density} is then quantified through the state-transfer fidelity
\begin{equation}
\mathcal F_{fi}(t)
=
\bra{\psi_{f\leftarrow i}}\rho_f(t)\ket{\psi_{f\leftarrow i}}.
\label{eq:state_transfer_fidelity}
\end{equation}
Using Eqs.~\eqref{eq:initial_qubit_density} and \eqref{eq:final_qubit_density}, one directly obtains
\begin{equation}
\begin{split}
\mathcal F_{fi}(t)
&=
|c_0|^2
+\left(|c_1|^4-|c_0|^2|c_1|^2\right)
\mathcal G_{fi}^{\mathrm{pop}}(t)\\
&+2|c_0|^2|c_1|^2
\operatorname{Re}\!\left[\mathcal G_{fi}^{\mathrm{coh}}(t)\right],
\end{split}
\label{eq:state_transfer_fidelity_explicit}
\end{equation}
where the population and coherence propagators follow Eq.~\eqref{eq:final_qubit_density}. 
Equation~\eqref{eq:state_transfer_fidelity_explicit} characterizes the transfer of one specific input qubit. To obtain a measure that does not depend on the particular state initially encoded, we average this fidelity over all pure qubit states. We therefore parameterize the coefficients appearing in Eq.~\eqref{eq:initial_qubit_state} as
\begin{equation}
c_0=\cos\frac{\theta}{2}
\quad \text{and} \quad
c_1=e^{i\phi}\sin\frac{\theta}{2},
\end{equation}
so that each pair $(\theta,\phi)$ defines one point on the Bloch sphere. 
The average fidelity is then obtained by averaging uniformly
over all pure qubit states on the Bloch sphere:
\begin{equation}
\overline{\mathcal F}_{fi}(t)
=
\frac{1}{4\pi}
\int_0^{2\pi}d\phi
\int_0^\pi d\theta\,\sin\theta\,
\mathcal F_{fi}(\theta,\phi,t).
\label{eq:bloch_average_fidelity}
\end{equation}
Hence, $\overline{\mathcal F}_{fi}(t)$ measures the ability of the system to transfer an arbitrary, a priori unknown, qubit state rather than a particular initial preparation. 
Using the Bloch-sphere averages $\langle |c_0|^2\rangle=1/2$, $\langle |c_1|^4\rangle=1/3$, and $\langle |c_0|^2|c_1|^2\rangle=1/6$, Eq.~\eqref{eq:bloch_average_fidelity} directly gives
\begin{equation}
\overline{\mathcal F}_{fi}(t)
=
\frac12
+\frac13\operatorname{Re}\!\left[\mathcal G_{fi}^{\mathrm{coh}}(t)\right]
+\frac16\mathcal G_{fi}^{\mathrm{pop}}(t).
\label{eq:average_transfer_fidelity}
\end{equation}
Following Ref.~\cite{Bose2003}, if we assume that the dynamical phase accumulated by the coherence propagator can be corrected locally at the final site, we have  $\operatorname{Re}[\mathcal G_{fi}^{\mathrm{coh}}(t)]$ replaced by $|\mathcal G_{fi}^{\mathrm{coh}}(t)|$ and therefore
\begin{equation}
\overline{\mathcal F}_{fi}(t)
=
\frac12
+\frac13\left|\mathcal G_{fi}^{\mathrm{coh}}(t)\right|
+\frac16\mathcal G_{fi}^{\mathrm{pop}}(t).
\label{eq:phase_corrected_average_fidelity}
\end{equation}
which is the form we employed in our simulations.
To conclude, note that in the absence of a phonon environment, the two propagators are related through $\mathcal G_{fi}^{\mathrm{pop}}(t)=|\mathcal G_{fi}^{\mathrm{coh}}(t)|^2$, and we recover the standard fidelity introduced by Bose~\cite{Bose2003}.

\section{Analytical Form of the Propagators\\ 
for One-Dimensional Coupled-State Blocks}
\label{app:nondegenerate_reduction} 

In this Appendix, we consider the particular limit where all
renormalized polaron--polariton states are nondegenerate and
effectively decoupled within the PB-VVPT description.
This corresponds to the limiting partition where the total
number of blocks is $N_{\mathcal B}=N+1$, with
$\dim(\mathcal B)=1$ for every coupled-state block.
In this case, the general many-body formulation introduced in
Sec.~\ref{sec:effective_propagators} admits an additional
analytical simplification that we detail here.

For a one-dimensional block
$\mathcal B=\{\ket{\widetilde\chi_\mu}\}$, the corresponding
effective Hamiltonian takes the form
\begin{equation}
 \widetilde H^{(\mathcal B)}
 =
 \ketbra{\widetilde\chi_\mu}{\widetilde\chi_\mu}
 \widetilde E_\mu
 +
 \ketbra{\widetilde\chi_\mu}{\widetilde\chi_\mu}
 \otimes
 \widetilde H_{\mathrm{ph}}^{(\mu)},
 \label{eq:conditional_effective_hamiltonian}
\end{equation}
where
\begin{equation}
 \widetilde H_{\mathrm{ph}}^{(\mu)}
 =
 H_{\mathrm{ph}}
 +\sum_{s,s'=1}^{N}
 \Lambda_{ss'}^{(\mu)}b_s^\dagger b_{s'},
 \ \text{ with } \
 \Lambda_{ss'}^{(\mu)}
 \equiv
 \widetilde\Lambda_{ss'}^{\mu\mu}.
 \label{eq:conditional_phonon_hamiltonian}
\end{equation}
The phonon dynamics therefore become conditional on the unique
polaron--polariton state $\ket{\widetilde\chi_\mu}$.

Since the matrix $\Lambda^{(\mu)}$ is Hermitian, the
corresponding conditional phonon Hamiltonian can be
diagonalized through a unitary transformation of the local
phonon modes.
To this end, we introduce the eigenvalues
$\delta\Omega_q^{(\mu)}$ and orthonormal eigenvectors
$\mathbf{U}_q^{(\mu)}$ of $\Lambda^{(\mu)}$, satisfying
\begin{equation}
 \sum_{s'=1}^{N}
 \Lambda_{ss'}^{(\mu)} U_{s'q}^{(\mu)}
 =
 \delta\Omega_q^{(\mu)} U_{sq}^{(\mu)},
 \label{eq:conditional_phonon_eigenvectors}
\end{equation}
where $U_{sq}^{(\mu)}$ denotes the $s$-th component
of the eigenvector associated with mode $q$.
The corresponding normal-mode creation and annihilation
operators are then defined as
\begin{equation}
 \widetilde b_q^{(\mu)}
 = \sum_{s=1}^{N} U_{sq}^{(\mu)*} b_s,
 \ \ \text{and} \ \
 \widetilde b_q^{(\mu)\dagger}
 = \sum_{s=1}^{N} U_{sq}^{(\mu)} b_s^\dagger.
 \label{eq:conditional_phonon_operators}
\end{equation}
This unitary transformation preserves the bosonic
commutation relations. Since all local phonon modes
have the same bare frequency $\Omega_0$, the free phonon
Hamiltonian remains invariant under this transformation.
The conditional phonon Hamiltonian therefore takes
the diagonal form
\begin{equation}
 \widetilde H_{\mathrm{ph}}^{(\mu)}
 =
 \sum_{q=1}^{N}
 \left(
 \Omega_0+\delta\Omega_q^{(\mu)}
 \right)
 \widetilde b_q^{(\mu)\dagger}
 \widetilde b_q^{(\mu)},
 \label{eq:diagonal_conditional_phonon_hamiltonian}
\end{equation}
where each $\delta\Omega_q^{(\mu)}$ describes the
phonon-frequency renormalization associated with
the polaron--polariton state
$\ket{\widetilde\chi_\mu}$.

The time evolution within the block then factorizes as
\begin{equation}
 e^{-i\widetilde H^{(\mathcal B)}t}
 =
 \ketbra{\widetilde\chi_\mu}{\widetilde\chi_\mu}
 e^{-i\widetilde E_\mu t}
 \otimes
 e^{-i\widetilde H_{\mathrm{ph}}^{(\mu)}t},
 \label{eq:conditional_propagator_factorization}
\end{equation}
where the propagator acts within the
corresponding one-dimensional polaron--polariton block.
As a consequence, the phonon dynamics can be evaluated without
introducing an explicit many-body phonon basis, and the
coherence and population propagators, originally defined in Eq.~\eqref{eq:block_population_fixed_n}, reduce to
\begin{equation}
 \begin{aligned}
 \mathcal G_{fi}^{\mathrm{coh}}(t)
 &=
 \sum_{\mu=1}^{N+1}
 \braket{f}{\widetilde\chi_\mu}
 \braket{\widetilde\chi_\mu}{i}
 e^{-i\widetilde E_\mu t}
 F^{(\mu)}(t),
 \\
 \mathcal G_{fi}^{\mathrm{pop}}(t)
 &=
 \sum_{\mu,\nu=1}^{N+1}
 \braket{f}{\widetilde\chi_\mu}
 \braket{\widetilde\chi_\mu}{i}
 \braket{i}{\widetilde\chi_\nu}
 \braket{\widetilde\chi_\nu}{f}\\
 &\qquad \qquad \qquad
 \times e^{-i(\widetilde E_\mu-\widetilde E_\nu)t}
 F^{(\mu\nu)}(t),
 \end{aligned}
 \label{eq:analytical_population_propagator}
\end{equation}
where the effect of the phonon environment is entirely
contained in the two thermal factors $F^{(\mu)}(t)$ and
$F^{(\mu\nu)}(t)$.

For the coherence propagator, the corresponding factor reads
\begin{equation}
 F^{(\mu)}(t)
 =
 \operatorname{Tr}_{\mathrm{ph}}
 \Bigl[
 \rho_{\mathrm{ph}}
 e^{iH_{\mathrm{ph}}t}
 e^{-i\widetilde H_{\mathrm{ph}}^{(\mu)}t}
 \Bigr]
 =
 \prod_{q=1}^{N}
 \frac{1-e^{-\beta\Omega_0}}
 {1-e^{-\beta\Omega_0-i\delta\Omega_q^{(\mu)}t}}.
 \label{eq:coherence_phonon_factor}
\end{equation}
This expression follows directly from the diagonal form of
$\widetilde H_{\mathrm{ph}}^{(\mu)}$ and reduces the thermal
trace to a product over the $N$ renormalized phonon modes.

For the population propagator, the interference between two renormalized
polaron--polariton states $\ket{\tilde \chi_\mu}$ and $\ket{\tilde \chi_\nu}$ involves instead
the product of two conditional phonon evolutions,
\begin{equation}
 \begin{aligned}
 F^{(\mu\nu)}(t)
 &=
 \operatorname{Tr}_{\mathrm{ph}}
 \Bigl[
 \rho_{\mathrm{ph}}
 e^{i\widetilde H_{\mathrm{ph}}^{(\nu)}t}
 e^{-i\widetilde H_{\mathrm{ph}}^{(\mu)}t}
 \Bigr]\\
 &=
 \frac{(1-e^{-\beta\Omega_0})^N}
 {\det\!\Bigl[
 I_N-e^{-\beta\Omega_0}
 e^{i\Lambda^{(\nu)}t}
 e^{-i\Lambda^{(\mu)}t}
 \Bigr]},
 \end{aligned}
 \label{eq:population_phonon_factor}
\end{equation}
where the second equality follows from the bosonic
trace-determinant identity detailed in Klich's work~\cite{Klich2003}.
In general, the ordered product of the two matrix
exponentials must be retained since
$\Lambda^{(\mu)}$ and $\Lambda^{(\nu)}$
do not necessarily commute.

Equations~\eqref{eq:analytical_population_propagator}--
\eqref{eq:population_phonon_factor}
provide closed analytical expressions for the PB-VVPT
propagators when all coupled-state blocks are one dimensional.
In this limit, the explicit many-body treatment of the phonon
Hilbert space (see Sec.~\ref{sec:numerical_evaluation}) is
completely avoided, and the numerical effort is reduced to
operations involving only $N\times N$ matrices.
More specifically, the diagonalization of the $N+1$
conditional phonon matrices $\Lambda^{(\mu)}$ requires
a computational cost of order $\mathcal O(N^4)$ using
dense linear algebra.
While the coherence propagator can then be evaluated directly
from the corresponding eigenvalues, the population propagator
involves matrix products and determinants for all pairs of
polaron--polariton states, leading to a cost of order
$\mathcal O(N^5)$ per time point in a straightforward
dense implementation.
Importantly, both calculations retain a polynomial scaling
with $N$, in contrast to the computational scaling derived
in Eq.~\eqref{eq:cost_effective_preparation} for the general
many-body PB-VVPT formulation, which explicitly depends
on the dimension of the truncated phonon Hilbert space.

\section{Convergence with Respect to the Phonon Cutoff}
\label{app:convergence}

In this appendix, we discuss the convergence of PB-VVPT and exact HTC calculations with respect to the maximum total phonon number $N_{\mathrm{ph}}^{\max}$.
We consider QST between opposite peripheral sites of the molecular wheel, $|i\rangle=|1\rangle$ and $|f\rangle=|3\rangle$, at $k_{\mathrm B}T/\Omega_0=0.25$, $0.50$, and $0.75$.
Parameters are $\lambda=8J$, $g=1J$, $\Omega_0=100J$, and $\omega_c=\varepsilon_0$, with $R_c=0.2$ for PB-VVPT.
Convergence is quantified through the maximum fidelity difference between successive cutoffs over the long-time interval considered:
\begin{equation}
\delta_{\mathrm{conv}}^{(N_{\mathrm{ph}}^{\max})}
= \max_{t\in[0,5000J^{-1}]}100
\left|
\overline{\mathcal F}_{fi}^{(N_{\mathrm{ph}}^{\max})}(t)
-
\overline{\mathcal F}_{fi}^{(N_{\mathrm{ph}}^{\max}-1)}(t)
\right|,
\label{eq:cutoff_convergence}
\end{equation}
which measures the sensitivity to the cutoff, expressed as a percentage.
We also monitor the omitted thermal weight
$1-Z_{\mathrm{ph}}^{\mathrm{trunc}}/Z_{\mathrm{ph}}$,
where $Z_{\mathrm{ph}}^{\mathrm{trunc}}$ is defined in
Eq.~\eqref{eq:Z_part} and
$Z_{\mathrm{ph}}=(1-e^{-\beta\Omega_0})^{-N}$.

\begin{figure}[h]
    \centering
    \includegraphics[width=\columnwidth]{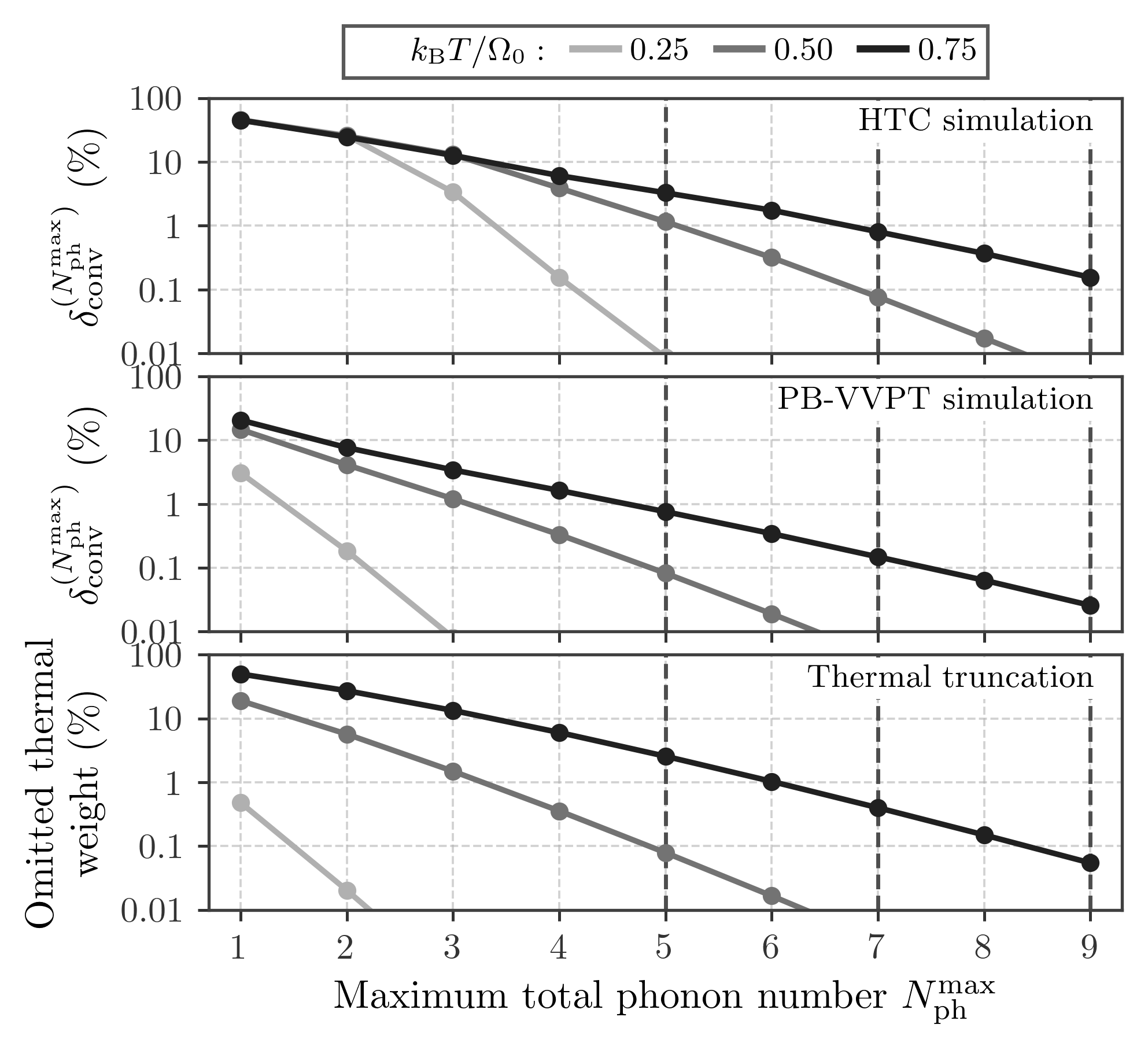}
    \caption{
    Convergence with respect to $N_{\mathrm{ph}}^{\max}$ for QST
    between opposite peripheral sites $|1\rangle$ and $|3\rangle$.
    Upper and middle panels: maximum fidelity differences between
    successive cutoffs over $t\in[0,5000J^{-1}]$ for exact HTC
    and PB-VVPT ($R_c=0.2$), respectively.
    Lower panel: omitted thermal weight, expressed as a percentage.
    The vertical dashed lines mark the cutoffs
    $N_{\mathrm{ph}}^{\max}=5$, $7$, and $9$ used in the benchmarks
    at $k_{\mathrm B}T/\Omega_0=0.25$, $0.50$, and $0.75$,
    respectively.
    Parameters are $\lambda=8J$, $g=1J$, $\Omega_0=100J$,
    and $\omega_c=\varepsilon_0$.
    }
    \label{fig:phonon_convergence}
\end{figure}

As shown in Fig.~\ref{fig:phonon_convergence}, both approaches
exhibit systematic convergence with increasing
$N_{\mathrm{ph}}^{\max}$, although higher temperatures require
larger cutoffs.
The values adopted in our benchmarks,
$N_{\mathrm{ph}}^{\max}=5$, $7$, and $9$ at the three respective
temperatures, are indicated by vertical dashed lines.
At these cutoffs, the HTC fidelity differences remain below
$0.2\%$ (upper panel), while PB-VVPT yields differences below
$0.03\%$ (middle panel).
The faster convergence of PB-VVPT reflects its fixed-phonon-number
structure, in which increasing the cutoff leaves the Hamiltonian
within each retained sector unchanged, unlike in HTC calculations
where different phonon sectors remain coupled.

Focusing now on the lower panel, we see that the omitted thermal
weight decreases as the cutoff $N_{\mathrm{ph}}^{\max}$ increases.
At higher temperatures, states with larger phonon occupations
become more populated, requiring higher cutoffs.
For the three cutoffs adopted at the respective temperatures,
the omitted thermal weight remains below $0.1\%$, supporting
their use in the numerical benchmarks.

\section{Optimal Light–Matter Coupling in the Molecular Wheel Network}
\label{app:Anlaytical_optimal_QST}

In this Appendix, we discuss the specific case of optimal matter-to-light QST from the core site to the cavity of the molecular wheel graph occurring in the absence of phonon environment when $g=J$.  

Let us disregard the coupling with the phonons and focus only on the exciton--cavity Hilbert space. 
In this case, the photon state $\ket{c}$, with energy $\omega_c$, interacts only with two excitonic states: the state localized at the core of the wheel, $\ket{0}$, with energy $\epsilon_0$, and the state uniformly delocalized over the $N_{p} =4$ peripheral sites, $\ket{\chi}$, with energy $\epsilon=\epsilon_0+J_{12}+J_{13}+J_{14}$. At resonance, $\omega_c=\epsilon_0$, the exciton-photon dynamics is therefore confined to a three-dimensional active subspace spanned by the basis $\{ \ket{c},\ket{0},\ket{\chi} \}$, in which the exciton-cavity Hamiltonian is given by
\begin{equation}
 H_{\mathrm{exc-cav}} 
 =
 \begin{pmatrix}
\epsilon_0 & g & \sqrt{N_p}g \\
g & \epsilon_0 & \sqrt{N_p}J \\
\sqrt{N_p}g & \sqrt{N_p}J & \epsilon
\end{pmatrix}
\end{equation}

Owing to the resonance condition $\omega_c=\epsilon_0$, we introduce a first set of polaritonic states defined as symmetric and antisymmetric superpositions of the excitonic state localized at the core and the photonic state:
\begin{equation}
\ket{\pi_{\pm}}=\frac{1}{\sqrt{2}}(\ket{0}\pm \ket{c})
\end{equation}
The two states $\ket{\pi_{\pm}}$ allow us to partially diagonalize the Hamiltonian. They are not coupled to each other and have energies $\epsilon_\pm=\epsilon_0\pm g$, respectively. However, a residual coupling with the peripheral excitonic state may still exist:
\begin{equation}
\bra{\chi}H\ket{\pi_{\pm}}=\sqrt{N_p/2}(J\pm g)
\end{equation}

In the intermediate basis ${\ket{\pi_-},\ket{\pi_+},\ket{\chi}}$, the Hamiltonian is a $3\times 3$ matrix that reads
\begin{equation}
 H_{\mathrm{exc-cav}}^{(3\times 3)}= \begin{pmatrix}
\epsilon_0-g & 0 & \sqrt{N_p/2}(J-g) \\
0 & \epsilon_0+g & \sqrt{N_p/2}(J+g) \\
\sqrt{N_p/2}(J-g) & \sqrt{N_p/2}(J+g) & \epsilon
\end{pmatrix}
\end{equation}

This expression reveals that, when $g=J$, the polaritonic state $\ket{\pi_-}$ becomes an eigenstate of the system. The two other eigenstates are superpositions of $\ket{\pi_+}$ and $\ket{\chi}$. They can be obtained by diagonalizing the following $2\times 2$ matrix:
\begin{equation}
 H_{\mathrm{exc-cav}}^{(2\times 2)}= \begin{pmatrix}
\epsilon_0+g & \sqrt{2N_p}g \\
\sqrt{2N_p}g & \epsilon
\end{pmatrix}
\end{equation}

To express the corresponding eigenstates in a compact form, we introduce the following parameters:
\begin{eqnarray}
\bar \epsilon &=&(\epsilon+\epsilon_+)/2 \nonumber \\
\Delta \epsilon&=&(\epsilon-\epsilon_+)/2 \nonumber \\
\tan(\gamma)&=&\sqrt{2N_p}g/\Delta \epsilon
\end{eqnarray}
where $\gamma$ belongs to $[0,\pi]$. The two polaritonic eigenstates are then given by:
\begin{eqnarray}
|\tilde\pi_+ \rangle&=&\sin(\gamma/2) \ket{\pi_+}+\cos(\gamma/2)\ket{\chi} \nonumber \\
|\tilde\pi_- \rangle&=&\cos(\gamma/2) \ket{\pi_+}-\sin(\gamma/2)\ket{\chi}
\end{eqnarray}
with the corresponding eigenvalues
\begin{equation}
\tilde \epsilon_\pm=\bar \epsilon \pm \sqrt{\Delta \epsilon^2+2N_pg^2}.
\end{equation}

In this context, matter-to-light QST from the central site of the molecular wheel, $\ket{i}=\ket{0}$, to the cavity mode, $\ket{f}=\ket{c}$, is entirely characterized by the transition amplitude $\bra{c}\exp(-iHt)\ket{0}$, which takes the following form in the polaritonic eigenbasis
\begin{equation}
\begin{split}
&\bra{c}\exp(-iHt)\ket{0} = \\
&-\frac{1}{2} e^{-i  \epsilon_-t}+ \frac{1}{2}\sin^2(\gamma/2) e^{-i\tilde\epsilon_+ t}+ \frac{1}{2}\cos^2(\gamma/2) e^{-i\tilde\epsilon_- t}.
\end{split}
\end{equation}
The triangle inequality provides an upper bound $\eta$ for the modulus of the propagator, defined as the sum of the moduli of the three probability amplitudes associated with the polaritonic pathways connecting $\ket{i}$ and $\ket{f}$. Over a sufficiently long timescale, the modulus of the propagator can approach this upper bound arbitrarily closely. Thus, $\eta$ provides an estimate of the maximum achievable transfer amplitude.

When $g=J$, we therefore obtain the following result:
\begin{equation}
\eta=\frac{1}{2}+ \frac{1}{2}\sin^2(\gamma/2) + \frac{1}{2}\cos^2(\gamma/2)=1.
\end{equation}
In other words, at resonance and provided that $g=J$, perfect QST is asymptotically achievable between the excitonic state localized at the core of the molecular wheel and the photonic state of the cavity.

\bibliography{mabiblio}

\end{document}